\documentclass[epjST]{svjour}
\usepackage{epsfig}
\usepackage{graphics,amsmath,amssymb}
\def\nA{nucleon-nucleus\ }
\def\pA{proton-nucleus\ }
\def\heA{$^3$He-nucleus\ }
\def\aA{$\alpha$-nucleus\ }
\def\AA{nucleus-nucleus\ }
\def\cx{charge-exchange\ }
\def\phe6{$^6$He+$p$\ }
\def\h6pn{$p(^6$He,$^6$Li$_{\rm IAS})n$\ }
\def\pn{$(p,n)$\ }
\def\he3t{$(^3$He,$t)$\ }
\def\xc{$^{14}$C$(^3$He,$t)^{14}$N\ }
\def\xca{$^{48}$Ca$(^3$He,$t)^{48}$Sc\ }

\def\pnZr{$^{90}$Zr$_{\rm g.s.}(p,n)^{90}$Nb$_{\rm IAS}$\ }

\def\pphe6{$p(^6$He,$^6$He)$p$\ }
\def\p2he6{$p(^6$He,$^6$He$^*)p$'\ }
\def\nli6{$^6$Li$^*+n$\ }
\def\An{$\tilde{a}+\tilde{A}$\ }
\def\Ap{$a+A$\ }

\def\pPb{$p+^{208}$Pb\ }
\def\nPb{$n+^{208}$Pb\ }

\def\bm#1{\mbox{\boldmath $#1$}}
\begin{document}
\title{Folding model study of the charge-exchange scattering to the 
 isobaric analog state and implication for the nuclear symmetry energy}
\author{Dao T. Khoa\inst{1}\fnmsep\thanks{\email{khoa@vinatom.gov.vn}}
\and Bui Minh Loc\inst{1,2} \and Dang Ngoc Thang\inst{1}}
\institute{Institute for Nuclear Science {\rm \&} Technique, VINATOM, 
 \\ 179 Hoang Quoc Viet Rd., Hanoi, Vietnam \and
 University of Pedagogy, Ho Chi Minh City, Vietnam}
\abstract{The Fermi transition ($\Delta L=\Delta S=0$ and $\Delta T=1$) 
between the nuclear isobaric analog states (IAS), induced by the \cx \pn 
or \he3t reaction, can be considered as ``elastic" scattering of proton 
or $^3$He by the isovector term of the optical potential (OP) that flips  
the projectile isospin. The accurately measured \pn or \he3t scattering 
cross-section to the IAS can be used, therefore, to probe the isospin
dependence of the proton or $^3$He optical potential. Within the folding
model, the isovector part of the OP is determined exclusively by the 
neutron-proton difference in the nuclear densities and the isospin 
dependence of the effective nucleon-nucleon (NN) interaction.    
Because the isovector coupling explicitly links the isovector part of
the proton or $^3$He optical potential to the cross section of the 
\cx \pn or \he3t scattering to the IAS, the isospin dependence 
of the effective (in-medium) NN interaction can be well tested in the 
folding model analysis of these charge-exchange reactions. On the other 
hand, the same isospin- and density dependent NN interaction can also 
be used in a Hartree-Fock calculation of asymmetric nuclear matter, 
to estimate the nuclear matter energy and its asymmetry part 
(the nuclear symmetry energy). As a result, the fine-tuning of the 
isospin dependence of the effective NN interaction against the measured 
\pn or \he3t cross sections should allow us to make some realistic 
prediction of the nuclear symmetry energy and its density dependence.}
\maketitle
\section{Introduction}
 \label{sec1}
The \cx \pn and \he3t reactions are the most suitable tool to excite 
the isobaric analog state (IAS) of the target nucleus. IAS has about the 
same structure as that of the target except for the replacement of a neutron 
by a proton and, hence, differs in energy approximately by the Coulomb 
energy of the added proton. Within the isospin symmetry, the two 
isobaric analog states are just two members of the isospin multiplet
which differ only in the orientation of the isospin $\bm{T}$. The similar
structures of the initial and final states of the \pn or \he3t reaction 
makes these reactions very much like the ``elastic" scattering in which the 
isospin of the incident proton or $^3$He is flipped \cite{Dri62,Sat64,Sat69}. 
In such a scenario for the \cx scattering to the IAS, the isospin-flip 
(elastic) scattering is naturally caused by the isovector term of the 
optical potential (OP) that is directly proportional to the neutron-proton 
asymmetry of the target nucleus, $\varepsilon=(N-Z)/A$. The empirical
isovector term of the \pA or \heA OP in the Woods-Saxon form has been 
used by Satchler {\it et al.} \cite{Dri62,Sat64} some 50 years ago 
as the \cx form factor to study the \cx \pn or \he3t scattering to the IAS 
within the distorted wave Born approximation (DWBA). 

Given the same isospin $t=1/2$ of nucleon and $^3$He, the central \nA 
or \heA OP for the elastic scattering on a nonzero-isospin target  
can be written in the following Lane form \cite{La62}
\begin{equation}
 U(R)=U_0(R)+4U_1(R)\frac{{\bm t}.{\bm T}}{aA}, \label{e1}
\end{equation}
where ${\bm t}$ is the isospin of the projectile and ${\bm T}$ 
is that of the target with mass number $A$, and $a=1$ and 3 for 
nucleon and $^3$He, respectively. The second term of Eq.~(\ref{e1}) is the 
\emph{symmetry term} of the OP, and $U_1$ is now known as the Lane 
potential that contributes to both the elastic scattering and \cx transition 
to the IAS \cite{Sat69}. The knowledge of $U_1$ is of fundamental interest 
for different studies of the nuclear phenomena in which neutrons and protons 
participate differently (isovector modes). In the \nA case, the relative 
contribution by the Lane potential $U_1$ to the elastic scattering cross section 
has been shown to be quite small and amounts only a few percent for a neutron-rich 
target \cite{Kho02,Kho03}. Nevertheless, the Fermi-type ($\Delta J^\pi=0^+,\ 
\Delta T=1$) transition strength of the \cx \pn reaction to the IAS is determined 
entirely by $U_1$. Therefore, the accurately measured data of the \pn scattering
to the IAS have been used successfully in the folding model analysis 
\cite{Kho05,Kho07} to probe the isospin dependent part of the folded \pA OP. 
It is complimentary to note that the volume and surface strengths of the symmetry 
term of the nuclear binding energy have been determined quite accurately 
from a systematic analysis of the excitation energies of the IAS \cite{Dan09}.

The nucleon OP has been studied over the years and there are several ``global" sets 
of the OP parameters deduced from the extensive optical model (OM) analyses of 
nucleon elastic scattering, like that by Becchetti and Greenlees \cite{BG69}, the 
CH89 global OP \cite{Va91}, and the systematics by Koning and Delaroche 
\cite{Kon03}. Although parametrized in the empirical Woods-Saxon form, these global 
systematics are very valuable in predicting the nucleon OP when elastic scattering 
data are not available or cannot be measured which is the case for the unstable, 
dripline nuclei. Given a large neutron excess in the unstable neutron-rich nuclei, 
it is important to know as accurate as possible the isospin dependence of the 
nucleon OP before using it in various studies of nuclear reactions and nuclear 
astrophysics. Because the high-quality \pn data are not available for a wide range 
of target masses and proton energies, the isovector term of the nucleon OP 
has been deduced \cite{BG69,Va91,Kon03} mainly from the OM studies of elastic proton 
and neutron scattering from the same target and at about the same energy, 
where the second term of Eq.~(\ref{e1}) has the same strength, but opposite 
signs for proton and neutron. Only in few cases the Lane potential $U_1$ has 
been deduced from the DWBA studies of the \cx \pn scattering to the IAS 
\cite{Car75,Jon00}.

Regarding the isospin dependence of the \heA OP, it has been very little
investigated. Even in a recent version of the \emph{global} OP for $^3$He 
and triton \cite{Pang09} the real OP contains no isovector term like that 
of Eq.~(\ref{e1}), and the purely isoscalar parametrization of the real OP
seems to deliver rather good OM description of the data, with a slight 
dependence on the neutron-proton asymmetry $\varepsilon$ of the surface
imaginary potential. Unlike the \nA case, the Lane consistency of the 
\heA OP has not been well established, and the measured data of the 
\cx \he3t scattering to the IAS have been mainly studied in the DWBA 
with the \cx form factor given by folding the
isospin-dependent part of an effective $^3$He-nucleon interaction with
the nuclear transition density for the IAS excitation \cite{Werf1,Werf2}. 
It is not straightforward, however, to link the isospin-dependence 
of the effective $^3$He-nucleon interaction to the isospin-dependence 
of the in-medium nucleon-nucleon (NN) interaction, the most vital input 
for a many-body study of asymmetric nuclear matter \cite{Kho07,Kho96}. 
In general, both the single-folding calculation of the nucleon OP 
\cite{Kho02,Kho07} and double-folding calculation of the \AA OP 
\cite{Kho96} give rise naturally to a non-zero isovector term 
of the OP (the microscopic prototype of the Lane potential) when both 
the projectile and target have non-zero isospins. Given the success 
of the single-folded nucleon OP in the description of the nucleon 
elastic scattering and \cx \pn scattering to the IAS \cite{Kho07}, 
the double-folded \heA OP is expected to give also a reasonable 
description of the elastic $^3$He scattering and \cx \he3t scattering 
to the IAS.    

Within the DWBA or coupled-channel (CC) analysis, the folded (Lane consistent) 
\nA or \heA OP serves as a direct link between the isospin dependence of the 
in-medium NN interaction and the \cx \pn or \he3t scattering to the IAS, so
that accurately measured \cx data can be used to probe the isospin dependence 
of the NN interaction. On the other hand, within a many-body calculation 
of nuclear matter (NM), the asymmetry of the equation of state (EOS) of
the NM depends entirely on the density- and isospin dependence of the in-medium 
NN interaction \cite{Kho96,Zuo99}. Such an asymmetry is determined by the 
\emph{nuclear symmetry energy} $S(\rho)$ defined in terms of the NM energy 
$E(\rho,\delta)$ as
\begin{equation}
E(\rho,\delta)=E(\rho,0)+S(\rho)\delta^2 
 \label{e2}
\end{equation}
where $\delta=(\rho_n-\rho_p)/\rho$ is the neutron-proton asymmetry parameter
(the NM limit of the neutron-proton asymmetry $\varepsilon$ in nuclei).
The knowledge about the nuclear EOS is well known to be vital for 
the understanding of the dynamics of supernova explosion and neutron star 
formation \cite{Bet90,Lat04,Ste05,Ba08}. The nuclear symmetry energy
determined at the saturation density, $E_{\rm sym}=S(\rho=\rho_0\approx 0.17$
fm$^{-3}$), is also known in the literature as the \emph{symmetry energy} 
or \emph{symmetry coefficient}. Although numerous nuclear many-body calculations 
have predicted $E_{\rm sym}$ to be around 30 MeV \cite{Kho96,Zuo99,Bra85,Pea00}, 
a direct experimental determination of $E_{\rm sym}$ still remains a challenge. 
Moreover, the knowledge about the density dependence of $S(\rho)$ is 
very important in modeling a realistic nuclear EOS and it has been, 
therefore, the main subject of many nuclear structure and reaction studies.
The main approach to probe $S(\rho)$ associated with a given 
in-medium NN interaction is to test this interaction in the simulation 
of heavy-ion (HI) collisions using transport and/or statistical models
\cite{Ba08,Tsa09,Da02,Ono03,Ch05,Ba09,She07,Sh07} or in the structure 
studies of nuclei with large neutron excess 
\cite{Dan09,Da03,Aru04,Tod05,Pie07,Pie09,Cen09,Tri08,Br00,Fur02}. 
Some conclusion on the low- and high-density behavior of $S(\rho)$ is then 
made based on the physics constraints implied by such studies. 

In about the same way, our recent folding model studies of the 
\pn scattering to the IAS \cite{Kho05,Kho07} were aimed to gain some 
information on the nuclear symmetry energy. The isospin dependence 
of the chosen in-medium NN interaction, fine-tuned to the 
best fit of the \pn data, has been shown \cite{Kho07} to give 
a symmetry coefficient $E_{\rm sym}$ very close to the empirical 
values deduced from other studies. However, the \pn data can be used 
to probe the isospin dependence of the in-medium NN interaction at 
sub-saturation nuclear densities only, i.e., $\rho<\rho_0$ where 
$\rho_0$ is the saturation density of the symmetric nuclear matter.    
Although similar in the isospin-coupling scheme, the high-quality data
of the \cx \he3t scattering to the IAS could be in general sensitive 
to higher nuclear densities ($\rho\gtrsim\rho_0$) formed in the spatial 
overlap of the $^3$He projectile with the target nucleus. This is 
an essential feature of the folding model analysis of elastic \AA 
scattering that the \AA OP at small internuclear radii is determined 
by the effective NN interaction at high nuclear medium densities
(see a recent review in Ref.~\cite{Kho07r}). 

In the present work, we discuss the results of a consistent folding 
model analysis of both the \pn and \he3t scattering to the IAS, 
which allowed us not only to validate the conclusion made earlier 
in the study of the \pn scattering \cite{Kho07} for the nuclear 
symmetry energy but also to probe the slope of $S(\rho)$ with the
increasing NM density.

\section{Isospin coupling and Lane equations for the \cx scattering 
to the IAS}
\label{sec2}
\subsection{General formalism}
We give here a brief introduction to the coupled-channel formalism for the
\cx \pn or \he3t scattering to the IAS, and the reader is
referred to Satchler's book \cite{Sat83} for more technical details. 
Let us consider a given isospin multiplet with fixed values of isospin 
$\bm{t}$ for the projectile and $\bm{T}$ for the target. Then, the 
isospin projections are $T_z=(N-Z)/2$ and $\tilde{T_z}=T_z-1$ for the 
target nucleus $A$ and \emph{isobaric analog nucleus} $\tilde{A}$, 
respectively, and the \cx transition between $A$ and $\tilde{A}$ induced 
by proton or $^3$He is like elastic scattering except for a reorientation 
of both isospins $\bm{t}$ and $\bm{T}$. We further denote, in the isospin 
representation, state formed by adding proton or $^3$He to $A$ as $|aA>$ 
and adding a neutron or triton to $\tilde{A}$ as $|\tilde{a}\tilde{A}>$. 
The transition matrix elements of the isovector part of the optical 
potential (\ref{e1}) for elastic scattering can be obtained \cite{Sat83} as
\begin{eqnarray}
 <aA|4U_1(R)\frac{{\bm t}.{\bm T}}{aA}|aA>&=&-\frac{2}{aA}T_zU_1(R),
 \nonumber\\
 <\tilde{a}\tilde{A}|4U_1(R)\frac{{\bm t}.{\bm T}}{aA}|\tilde{a}\tilde{A}>
 &=&\frac{2}{aA}(T_z-1)U_1(R). \label{e3a}
\end{eqnarray}
where $\bm{R}$ is the radial separation between the projectile and
target. Similarly, the transition matrix element or \cx form factor 
(FF) for the \cx \pn or \he3t scattering to the IAS is obtained as
\begin{equation}
 <\tilde{a}\tilde{A}|4U_1(R)\frac{{\bm t}.{\bm T}}{aA}|aA>\equiv
 F_{\rm cx}(R)=\frac{2}{aA}\sqrt{2T_z}U_1(R).
 \label{e3b}
\end{equation}
In the two-channel approximation for the \cx scattering to the IAS,
the total wave function is written as
\begin{equation}
\Psi=|aA>\chi_{aA}({\bm R})+|\tilde{a}\tilde{A}>
 \chi_{\tilde{a}\tilde{A}}({\bm R}),
 \label{e4}
\end{equation}
where the waves $\chi({\bm R})$ describe the relative motion 
of the scattering system. Then, the elastic and \cx scattering 
cross sections are readily obtained from the solutions of the 
following (coupled-channel) Lane equations \cite{Sat83}
\begin{eqnarray}
\left[K_a+U_a(R)-E_a\right]
 \chi_{aA}({\bm R})=-F_{\rm cx}(R)\chi_{\tilde{a}\tilde{A}}({\bm R}),
  \label{e5a} \\
\left[K_{\tilde{a}}+U_{\tilde{a}}(R)-E_{\tilde{a}}\right]
 \chi_{\tilde{a}\tilde{A}}({\bm R})=-F_{\rm cx}(R)\chi_{aA}({\bm R}).
 \label{e5b}
\end{eqnarray}
Here $K_{a(\tilde{a})}$ and $E_{a(\tilde{a})}$ are the kinetic-energy 
operators and center-of-mass energies of the \Ap and \An partitions. 
The OP in the entrance (\Ap) and exit (\An) channels are determined 
explicitly through the isoscalar ($U_0$) and isovector ($U_1$) parts 
of the optical potential (\ref{e1}) as
\begin{eqnarray}
U_a(R)& = & U_0(R)-\frac{2}{aA}T_zU_1(R), \label{e6a} \\
U_{\tilde{a}}(R)& = & U_0(R)+\frac{2}{aA}(T_z-1)U_1(R). \label{e6b}
\end{eqnarray}
In the actual CC calculation, $U_a$ and $U_{\tilde{a}}$ are added 
by the corresponding spin-orbital and Coulomb potentials (the Coulomb 
term in the exit channel is nonzero only if $\tilde{a}$ is triton). 
Since the energies of the isobar analog states are separated 
approximately by the Coulomb displacement energy, the \cx transition 
between them has a nonzero $Q$ value. To account for this effect, 
the isoscalar $U_0$ and isovector $U_1$ potentials used to construct 
$F_{\rm cx}(R)$ are evaluated from the proton or $^3$He optical potential
at the energy of $E=E_{\rm lab}-Q/2$ \cite{Sat64}, and those used to 
construct $U_{\tilde{a}}(R)$ are evaluated from the neutron or triton OP 
at the energy $E=E_{\rm lab}-Q$.

For the \pn scattering to the IAS, the existing global parameters 
for the nucleon OP \cite{BG69,Va91,Kon03} can be used to construct 
the isoscalar $U_0$ and isovector $U_1$ parts of the OP used 
in the Lane equations (\ref{e5a})-(\ref{e5b}) \cite{Kho07}. 
Unlike the \nA case, the isospin dependence of the \heA OP has been 
investigated in terms of an effective $^3$He-nucleon interaction
only \cite{Werf1,Werf2}. Moreover, the existing elastic $^3$He or 
triton scattering data were shown to be well reproduced by a 
(phenomenological) global OP \cite{Pang09}, whose real part is 
purely isoscalar and imaginary part contains a weak dependence 
on the asymmetry parameter $\varepsilon$ at the surface. Therefore, 
the folding model is probably the only consistent microscopic approach 
to evaluate FF using the same in-medium NN interaction for both \pn and 
\he3t \cx scattering to the IAS.

\subsection{Folding model}
In general, the central \nA or \AA potential $U$ is evaluated 
by the single-folding \cite{Kho02,Kho03} or double-folding 
\cite{Kho96,Kho07r,Kho97,Kho00} approach as the following 
Hartree-Fock-type potential
\begin{equation}
  U=\sum[<ij|v_{\rm D}|ij>+<ij|v_{\rm EX}|ji>], \label{e7}
\end{equation}
where single-folding or double-folding summation is performed over all 
nucleon states of the target ($j\in A$) or of both the target and 
projectile ($i\in a,j\in A$), respectively.  
$v_{\rm D}$ and $v_{\rm EX}$ are the direct and exchange parts of the
(effective) NN interaction between the projectile nucleon $i$ and target
nucleon $j$. The antisymmetrization of the \nA or \AA system is done by
taking into account the knock-on exchange effects. To separate the 
isovector part of $U$ which gives rise to the Lane potential, one needs 
to make explicit the isospin degrees of freedom. Namely, the following 
spin-isospin decomposition of the (energy- and density dependent) NN 
interaction is used
\begin{eqnarray}
v_{\rm D(EX)}(E,\rho,s)=v^{\rm D(EX)}_{00}(E,\rho,s)+
 v^{\rm D(EX)}_{10}(E,\rho,s)(\bm{\sigma\sigma}') \nonumber\\
  +  v^{\rm D(EX)}_{01}(E,\rho,s)(\bm{\tau\tau}')+
 v^{\rm D(EX)}_{11}(E,\rho,s)(\bm{\sigma\sigma}')(\bm{\tau\tau}'),
\label{e8}
\end{eqnarray}
where $s$ is the internucleon distance and $\rho$ is the nuclear density around
the interacting nucleon pair. The contribution from the spin dependent terms
($v_{10}$ and $v_{11}$) in Eq.~(\ref{e8}) to the central \nA or \AA potential
(\ref{e7}) is exactly zero for a spin-saturated target like those considered in
the present work.

Using the explicit proton ($\rho_p$) and neutron ($\rho_n$) densities in the
folding input, the \nA or \AA optical potential (\ref{e7}) can be obtained 
explicitly \cite{Kho02,Kho96} in terms of the isoscalar ($U_{\rm IS}$) and 
isovector ($U_{\rm IV}$) parts as
\begin{equation}
 U(E,\bm{R})=U_{\rm IS}(E,\bm{R})\pm U_{\rm IV}(E,\bm{R}),
\label{e9}
\end{equation}
where - sign pertains to the proton or $^3$He optical potential (used 
for the entrance channel) and + sign to the neutron or triton OP (used 
for the exit channel). Each term in Eq.~(\ref{e9}) consists of the 
corresponding direct and exchange potentials. 

In the \nA case, the isoscalar and isovector potentials are given
by the single-folding approach \cite{Kho02} as 
\begin{eqnarray}
 U_{\rm IS}(E,\bm{R})=\int\{[\rho_n(\bm{r})+\rho_p(\bm{r})]
 v^{\rm D}_{00}(E,\rho,s) \nonumber\\
  +  [\rho_n(\bm{R},\bm{r})+\rho_p(\bm{R},\bm{r})]
 v^{\rm EX}_{00}(E,\rho,s)j_0(k(E,R)s)\}d^3r,
\label{e10}
\end{eqnarray}
\begin{eqnarray}
 U_{\rm IV}(E,\bm{R})=\int\{[\rho_n(\bm{r})-\rho_p(\bm{r})]
 v^{\rm D}_{01}(E,\rho,s) \nonumber\\
  +  [\rho_n(\bm{R},\bm{r})-\rho_p(\bm{R},\bm{r})]
 v^{\rm EX}_{01}(E,\rho,s)j_0(k(R)s)\}d^3r, \label{e11}
\end{eqnarray}
where $\bm{s}=\bm{r}-\bm{R},\ \rho(\bm{r},\bm{r}')$ is the one-body 
density matrix of the target, with $\rho(\bm{r})\equiv\rho(\bm{r},\bm{r}),
\ j_0(x)$ is the zero-order spherical Bessel function, and the local
relative motion momentum $k(R)$ is determined from
\begin{equation}
 k^2(E,R)=\frac{2\mu}{{\hbar}^2}[E_{\rm c.m.}-V(E,R)-V_C(R)].
\label{e12}
\end{equation}
Here, $\mu$ is the reduced mass, $V(E,R)$ and $V_C(R)$ are,
respectively, the real central nuclear and Coulomb parts of the
OP ($V_C(R)=0$ in the neutron-nucleus case). More details 
of the single-folding approach to evaluate $U_{\rm IS}$ and 
$U_{\rm IV}$ can be found in Ref.~\cite{Kho02}. If the effective
NN interaction is complex then $U_{\rm IS(IV)}$ should be treated
explicitly in terms of the real $V_{\rm IS(IV)}$ and imaginary
$W_{\rm IS(IV)}$ parts as
\begin{equation}
 U_{\rm IS(IV)}(E,\bm{R})=V_{\rm IS(IV)}(E,\bm{R})+
 iW_{\rm IS(IV)}(E,\bm{R}) \nonumber\
\end{equation}

In the \heA or triton-nucleus case, the isoscalar and isovector 
potentials are given in a similar manner by the double-folding 
approach \cite{Kho96} as 
\begin{eqnarray}
 U_{\rm IS}(E,\bm{R})=\int\int[\rho_1(\bm{r}_1)\rho_2(\bm{r}_2)
 v^{\rm D}_{00}(E,\rho,s)+\rho_1(\bm{r}_1,\bm{r}_1+\bm{s}) \nonumber \\
 \times\rho_2(\bm{r}_2,\bm{r}_2-\bm{s})
 v^{\rm EX}_{00}(E,\rho,s)j_0(k(E,R)s/M)]d^3r_1d^3r_2,
\label{e13}
\end{eqnarray}
\begin{eqnarray}
 U_{\rm IV}(E,\bm{R})=\int\int[\Delta\rho_1(\bm{r}_1)\Delta\rho_2(\bm{r}_2)
 v^{\rm D}_{01}(E,\rho,s)+\Delta\rho_1(\bm{r}_1,\bm{r}_1+\bm{s}) \nonumber \\
 \times\Delta\rho_2(\bm{r}_2,\bm{r}_2-\bm{s})
 v^{\rm EX}_{01}(E,\rho,s)j_0(k(E,R)s/M)]d^3r_1d^3r_2.\label{e14} 
\end{eqnarray}
Here, $\rho_i=\rho_{in}+\rho_{ip}\ {\rm and}\ 
\Delta\rho_i=\rho_{in}-\rho_{ip},\ \bm{s}=\bm{r}_2-\bm{r}_1+\bm{R}\ $ 
and $M=aA/(a+A)$. The local relative motion momentum $k(E,R)$ is determined
by the same formula (\ref{e12}), but with $\mu,\ V(E,R)$ and $V_{\rm C}(R)$ 
determined consistently for the $^3$He or triton OP. 

The \cx FF for both the \pn and \he3t scattering to the IAS is readily 
obtained in terms of the folded isovector potential $U_{\rm IV}$ as  
\begin{equation}
  F_{\rm cx}(R)=\frac{2}{aA}\sqrt{2T_z}U_1(R)=\sqrt{\frac{2}{T_z}}
  U_{\rm IV}(R) \label{e15}
\end{equation}

All the optical model (OM) calculation of elastic scattering and CC 
calculation of the \cx scattering to the IAS, with the folded OP and 
\cx FF, were done using external-potential option of the CC code 
ECIS97 written by Raynal \cite{Ra97}.

Given the isovector folded potentials (\ref{e11}) and (\ref{e14}) 
determined entirely by the difference $\Delta\rho$ between the neutron
and proton densities, it is necessary to have the nuclear densities 
determined as accurate as possible for a realistic folding model 
prediction of the \cx form factor (\ref{e15}). In the present work, 
we have used for the $^{90}$Zr and $^{208}$Pb targets the empirical 
neutron and proton densities deduced from the high-precision elastic 
proton scattering at 800 MeV by Ray {\it et al.} \cite{Ray78}. For   
the $^{48}$Ca and $^{120}$Sn targets the microscopic nuclear densities given 
by the Hartree-Fock-Bogoljubov approach \cite{Gr01} have been used. We have 
used  neutron and proton densities of $^{14}$C target given by the independent 
particle model \cite{Sa80}, which generates realistic wave function for each 
single-particle orbital using an appropriate Woods-Saxon 
potential for the bound state problem. The neutron and proton densities of $^3$He 
and triton given by the microscopic three-body calculation \cite{Nie01} using 
Argonne NN potential have been used in the double-folding calculation 
(\ref{e13}) and (\ref{e14}).   

\subsection{Isospin- and density dependent CDM3Y6 interaction}

Together with the nuclear densities, an appropriately chosen effective 
(energy- and density dependent) NN interaction $v^{\rm D(EX)}_{00(01)}$ 
is the most important input for the folding calculations 
(\ref{e10})-(\ref{e11}) and (\ref{e13})-(\ref{e14}). 
We have used in the present study the density dependent CDM3Y6 interaction 
\cite{Kho97} that is based on the M3Y interaction deduced from the G-matrix 
elements of the Paris NN potential \cite{Ana83}. The density dependence of the 
\emph{isoscalar} (IS) part of the CDM3Y6 interaction was introduced earlier in 
Ref.~\cite{Kho97} and its parameters have been carefully tested in numerous 
folding model analyses \cite{Kho07r,Kho97,Kho00,Kho95,Kho01} of the elastic, 
refractive \AA and \aA scattering. Because the isospin dependent term $v_{01}$ 
of the effective NN interaction can be directly probed in a folding model 
analysis of the \cx reaction, we have developed recently \cite{Kho07} 
an accurate procedure to parametrize the \emph{isovector} (IV) density dependence 
of the CDM3Y6 interaction based on the Brueckner-Hartree-Fock (BHF) results 
for the energy- and density dependent nucleon OP in nuclear matter by 
Jeukenne, Lejeune and Mahaux (JLM) \cite{Je77}. 

We recall that the isoscalar density dependence of the CDM3Y6
interaction has been introduced \cite{Kho97} as
\begin{eqnarray}
 v^{\rm D(EX)}_{00}(E,\rho,s)&=& F_{\rm IS}(E,\rho)v^{\rm D(EX)}_{00}(s),
\label{g1} \\ {\rm where}\ 
 F_{\rm IS}(E,\rho)&=& g(E)C_0[1+\alpha_0\exp(-\beta_0\rho)-\gamma_0\rho].
\label{g2}
\end{eqnarray}
Parameters of $F_{\rm IS}(\rho)$ were chosen \cite{Kho97} to reproduce the 
empirical saturation energy and density of the symmetric NM with an 
incompressibility $K\approx 252$ MeV, in the Hartree-Fock (HF) approximation. 
These parameters as well as those corresponding to other $K$ values can be 
found in Ref.~\cite{Kho07r}. The linear factor $g(E)\approx 1-0.0026E$ accounts 
effectively for the energy dependence of the IS density dependence, 
where $E$ is the incident energy per nucleon. Given the success of the 
parametrization (\ref{g1})-(\ref{g2}) in numerous folding model analyses 
of \nA and \AA scattering, a similar functional has been adopted 
\cite{Kho07} for the IV density dependence of the CDM3Y6 interaction
\begin{eqnarray}
 v^{\rm D(EX)}_{01}(E,\rho,s)&=& F_{\rm IV}(E,\rho)v^{\rm D(EX)}_{01}(s),
\label{g3} \\ {\rm where}\ 
 F_{\rm IV}(E,\rho)&=& C_1[1+\alpha_1\exp(-\beta_1\rho)-\gamma_1\rho].
\label{g4}
\end{eqnarray}
Instead of implying a simple linear energy dependence like $g(E)$, parameters 
of $F_{\rm IV}(E,\rho)$ were adjusted carefully at each considered energy $E$ 
to reproduce in the HF approximation the microscopic BHF results for the 
nucleon OP in nuclear matter by JLM group \cite{Je77}. The radial part 
of the IS and IV interactions, $v^{\rm D(EX)}_{00(01)}(s)$, were 
kept unchanged as derived in terms of three Yukawas \cite{Kho96} from 
the M3Y-Paris interaction \cite{Ana83}. 
\begin{figure}[bht] \vspace*{0cm}\hspace{0.5cm}
\includegraphics[width=\textwidth]{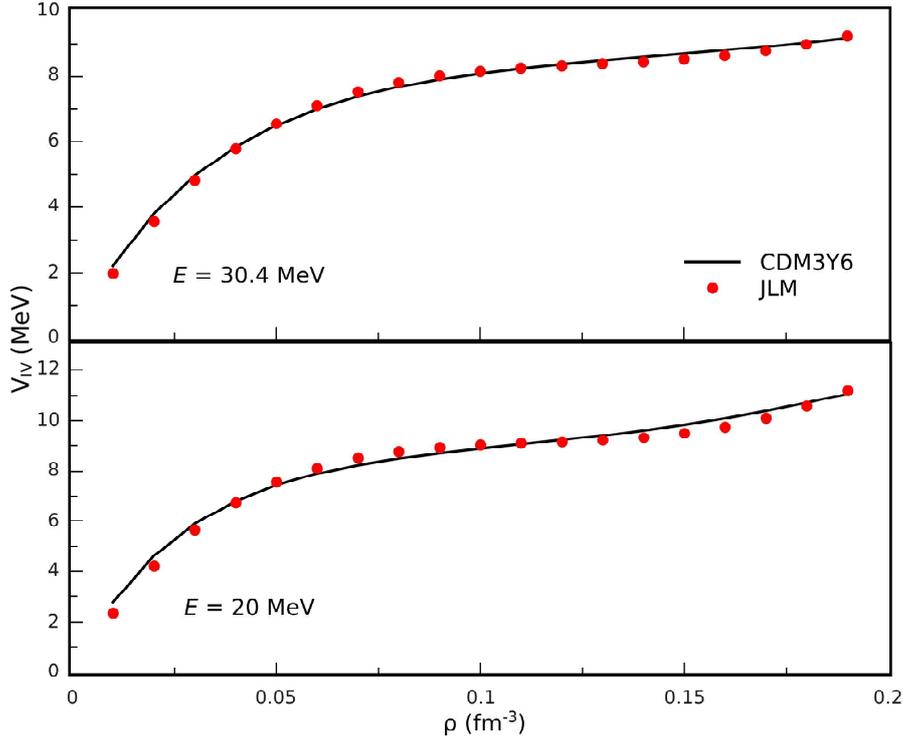}\vspace*{0.5cm}
 \caption{Real isovector nucleon optical potential $V_{\rm IV}(E,\rho)$ 
in NM (solid curves) given by the HF calculation \cite{Kho07} using the 
isovector CDM3Y6 interaction (\ref{g3})-(\ref{g4}) adjusted to reproduce
the JLM results (circles) \cite{Je77} at $E=30.4$ and 20 MeV.}  
 \label{f1a}
\end{figure}
\begin{figure}[bht] \vspace*{0cm}\hspace{0.5cm}
\includegraphics[width=\textwidth]{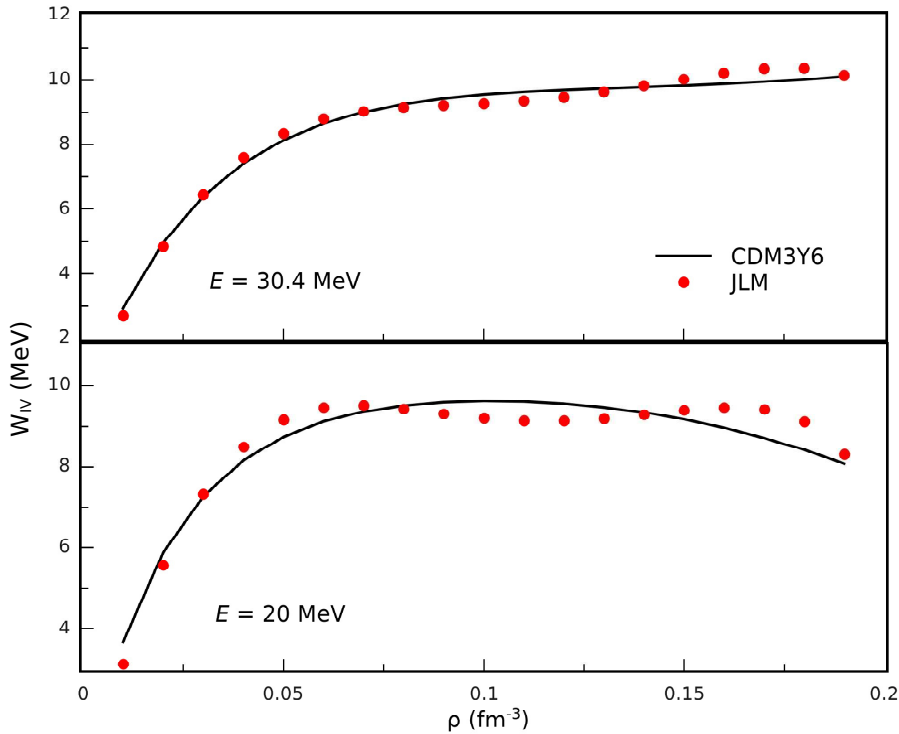}\vspace*{0.5cm}
 \caption{The same as Fig.~\ref{f1a} but for the imaginary 
potential $W_{\rm IV}(E,\rho)$.}  
 \label{f1b}
\end{figure}
Since the original M3Y interaction $v^{\rm D(EX)}_{00(01)}(s)$ is real, 
we have generated in a similar manner the imaginary parts of 
$F_{\rm IS(IV)}(E,\rho)$ at each considered energy $E$ based on the  
imaginary part of the JLM nucleon OP (see details in Ref.~\cite{Kho07}), 
so that the \emph{complex} density dependent CDM3Y6 interaction 
(\ref{g1})-(\ref{g4}) can be used in the folding model to predict 
both the real and imaginary parts of the OP. The fine tuning of the 
(isovector) density dependent parameters to the JLM results has been 
done carefully at each energy $E$ as shown in Figs.~\ref{f1a}-\ref{f1b}.
As a result, the \emph{complex} isovector density dependence of the 
present CDM3Y6 interaction (\ref{g3})-(\ref{g4}) is fully based on 
the JLM results for the complex isovector nucleon OP in NM \cite{Je77}.  

\section{Results and discussions}
\label{sec3}
\subsection{Folding model analysis of the \pn scattering to the IAS}
To study the \cx\pn scattering to the IAS based on the CC equations 
(\ref{e5a})-(\ref{e5b}), one needs to determine the nucleon OP in the entrance 
($U_a=U_p$) and exit ($U_{\tilde{a}}=U_n$) channels as accurate as possible. 
In general, the elastic neutron scattering on a nucleus being in the 
\emph{excited} IAS cannot be measured because most of isobar analog states 
are either a short-lived bound state or an unbound resonance. We have determined, 
therefore, $U_{\tilde{a}}$ from the isoscalar $U_0$ and isovector $U_1$ parts 
of the neutron OP evaluated at the energy $E=E_{\rm lab}-Q$, using Eq.~(\ref{e6b}). 

The phenomenological \nA global optical potentials \cite{BG69,Va91,Kon03} 
have been carefully determined based on large experimental databases of both the 
elastic \nA scattering and analyzing power angular distributions. It is helpful, 
therefore, to use them as the reference potentials in the present study. 
It should be noted that the isovector strength of the \nA OP is usually 
about 2-3\% of the total OP and its contribution to the elastic scattering 
cross section is too weak to allow us to probe the isospin dependence of the 
OP directly in the OM analysis of elastic scattering. 
Consequently, in a ``Lane consistent" approach, the isospin dependence 
of the \nA OP can be probed either in a OM study of the proton and neutron
elastic scattering at the appropriate energies \cite{deVito} or in the CC 
analysis of the \cx \pn scattering to the IAS \cite{Kho07}. In the latter case, 
the \cx form factor (\ref{e3b}) used in the CC equations (\ref{e5a})-(\ref{e5b}) 
is determined entirely by the Lane potential $U_1$, and the accurately measured 
\pn cross section can be used to fine tune the $U_1$ strength.   
Although the isospin dependence of the mentioned global nucleon OP's has not been 
calibrated against the DWBA or CC description of \cx \pn scattering to the IAS, 
the two more recent global OP's \cite{Va91,Kon03} were shown \cite{Kho07} 
to give a rather good description of the \pn data measured with $^{48}$Ca, 
$^{90}$Zr, $^{120}$Sn, and $^{208}$Pb targets at around 40 MeV \cite{Doe75}, 
with a slightly better fit to the data given by the CH89 global OP by 
Varner {\it et al.} \cite{Va91}. 

\begin{figure}[bht] \vspace*{0cm}\hspace{0.5cm}
\includegraphics[width=\textwidth]{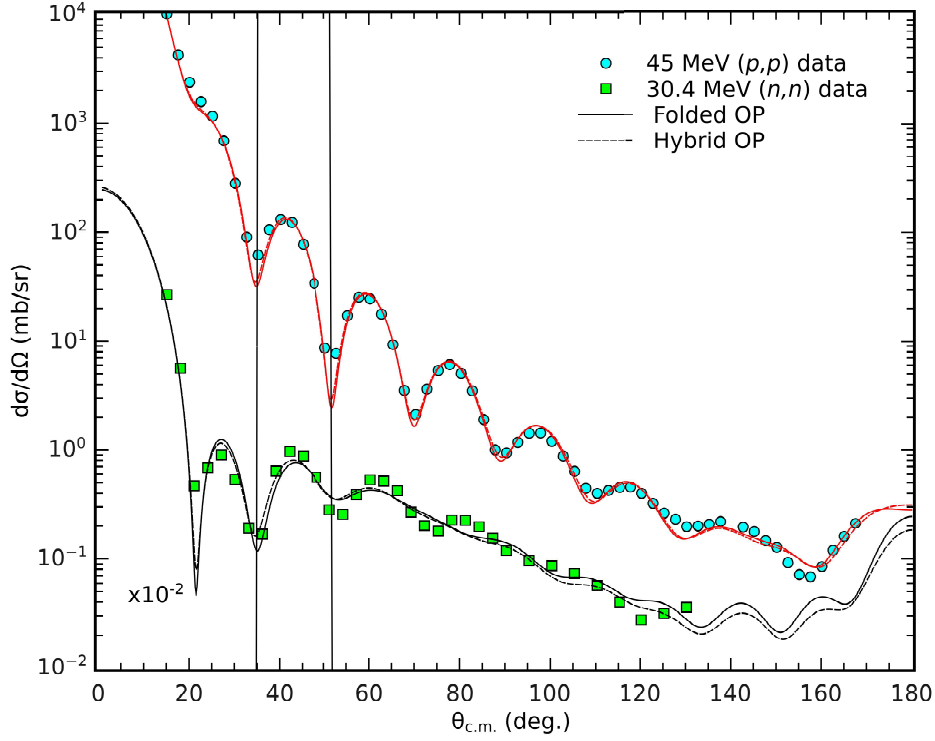}\vspace*{0.5cm}
\caption{OM description of the elastic \nPb data at 30.4 MeV
\cite{deVito} and \pPb data at 45 MeV \cite{vanO74} given by the 
complex folded OP and hybrid (real folded + imag. WS) OP.} 
 \label{f1}
\end{figure}

We recall further that the isospin is not a good quantum number in the 
repulsive Coulomb field of the nucleus that slows down the incident proton. 
It is necessary, therefore, to add the Coulomb correction 
$\Delta E_C$ to the incident proton energy and $\Delta U_C$ to $U_p$ 
to separate the main effects of the Coulomb field \cite{deVito} and 
restore the Lane consistency for the remainder of the OP. Estimation 
of the isospin impurity due to the Coulomb correction is not straightforward 
\cite{deVito}. One need first to determine correction $\Delta E_C$ to the 
incident proton energy such that the same isoscalar and isovector potentials 
$U_{0(1)}$ can be used to generate the proton and neutron OP at the energy 
$E_p$ and $E_n=E_p-\Delta E_C$, respectively. The Coulomb correction 
to the CH89 global OP for the \pPb system at 45 MeV has been determined 
recently \cite{deVito}, based on a comparative OM study of the proton 
and neutron elastic scattering from the lead target. In this case, 
$\Delta E_C$ was found to be about 14.6 MeV, which is lower than the 
empirical (energy-independent) value $\Delta E_C=6Ze^2/(5R_C)\approx 19$ MeV 
given by the original CH89 systematics \cite{Va91}. Using the 
prescription of Ref.~\cite{deVito} to relate the bombarding energies 
$E_p$ and $E_n=E_p-\Delta E_C$ so that the diffraction maxima and minima 
of the elastic proton and neutron angular distributions fall at about the 
same angles in the forward region (see Fig.~\ref{f1}), we have estimated 
that $\Delta E_C\approx 15$ MeV for the elastic \pPb scattering at 35 MeV. 
For other targets the effect caused by the Coulomb correction is weaker and 
we have used for simplicity the empirical $\Delta E_C$ value given 
by the CH89 global OP. 

In the folding model, the effect of the Coulomb potential to the
proton OP is taken into account self-consistently via the local relative 
motion momentum (\ref{e12}), and one needs only to use $E=E_p-\Delta E_C$ 
in the energy dependent factor $g(E)$ of the real IS density dependence. 
The parameters of Im$F_{\rm IS}(E,\rho)$ as well as those of both the 
real and imaginary parts of $F_{\rm IV}(E,\rho)$ must
be adjusted to the JLM results for the nucleon OP at the energy 
$E=E_p-\Delta E_C$. For example, the density dependent parameters 
used to calculate the \pPb optical potential at the proton incident 
energies of 35 and 45 MeV have been adjusted to the JLM results (see 
Figs.~\ref{f1a}-\ref{f1b}) at $E=20$ and 30.4 MeV, respectively. Thus, 
the proton OP of the entrance channel (\ref{e6a}) is given by the folding 
calculation using the complex CDM3Y6 interaction determined at the (Coulomb 
corrected) incident energy $E=E_p-\Delta E_C$. Similarly, the \cx form factor 
(\ref{e15}) of the \pn scattering to the IAS is given by the IV part 
of the folded proton OP obtained with the CDM3Y6 interaction at  
$E=E_p-\Delta E_C-Q/2$. The neutron OP of the exit channel    
(\ref{e6b}) is given by the folding calculation using the CDM3Y6 
interaction determined at the energy of emitting neutron $E=E_p-Q$.   

\begin{figure}[bht] \vspace*{-1cm}\hspace{-1cm}
\includegraphics[width=1.2\textwidth]{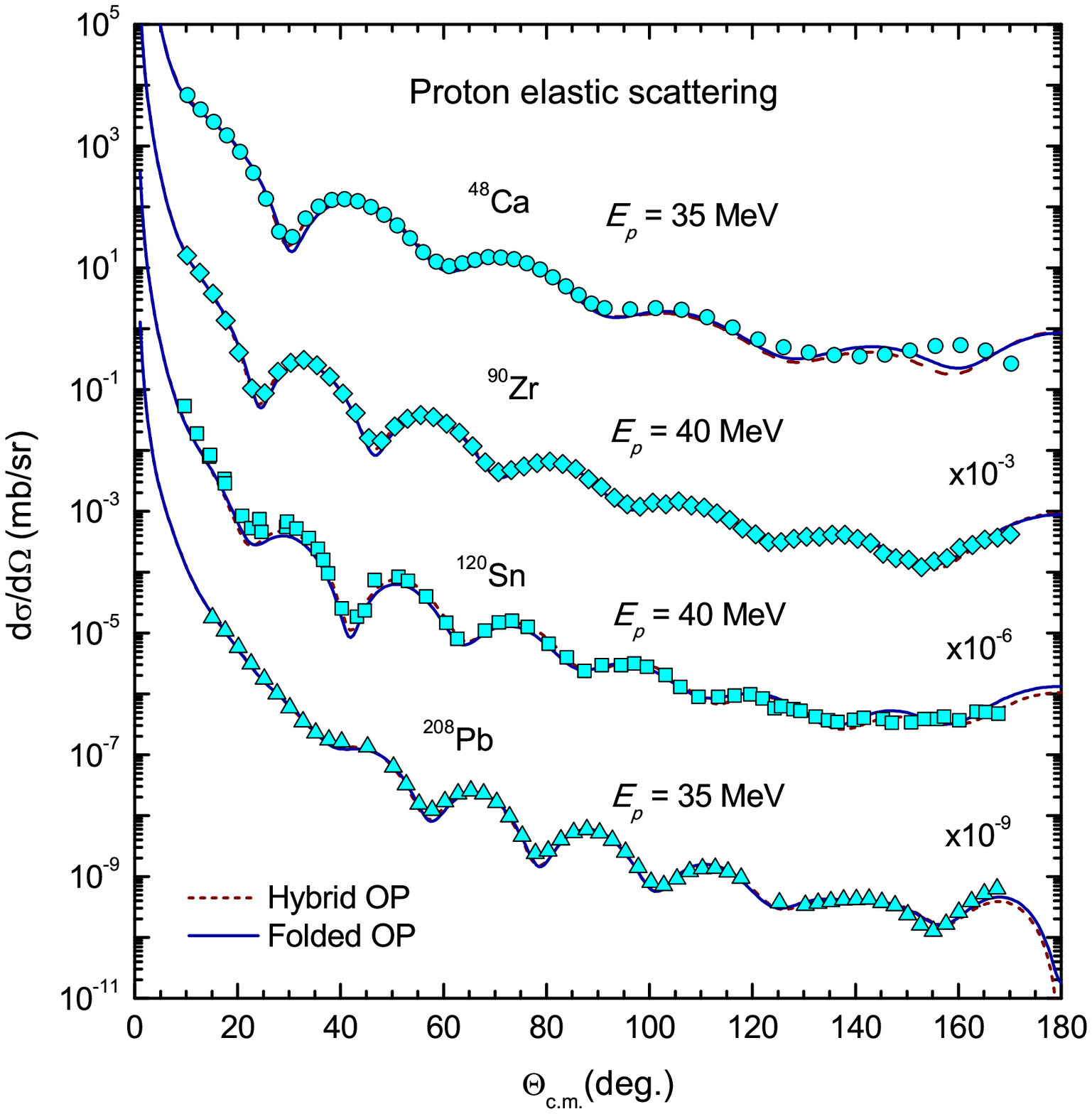}\vspace*{-1cm}
 \caption{OM description of the elastic proton scattering from $^{48}$Ca,
$^{90}$Zr, $^{120}$Sn, and $^{208}$Pb targets at 35 and 40 MeV given by the
complex folded OP (\ref{g5}) and hybrid OP (\ref{g6}). The data were taken
from Refs.~\cite{vanO74,Camis86,Fri67,Blu66}.}  
 \label{f4}
\end{figure}
Given the complex density dependence of the CDM3Y6 interaction 
appropriately determined at each energy, it is natural to check the 
OM description of the elastic proton scattering at the considered
energies using the complex folded OP (given by the CDM3Y6 interaction
determined at $E=E_p-\Delta E_C$)
\begin{equation}
 U(R)=N_V[V_{\rm IS}(R)-V_{\rm IV}(R)]+ 
 iN_W[W_{\rm IS}(R)-W_{\rm IV}(R)].
\label{g5}
\end{equation}
$U$ is further added by the spin-orbital and the Coulomb potential taken, 
for simplicity, from the CH89 systematics \cite{Va91}. The strengths 
$N_{V(W)}$ of the complex folded OP are adjusted to the best OM fit to 
the elastic scattering data. 

Because a high accuracy of distorted waves is always needed for the 
DWBA or CC calculation of the \cx scattering, we have used also a 
hybrid choice of the complex OP with the real part given by the 
folding model and imaginary part given by a Woods-Saxon (WS) potential
\begin{eqnarray}
 U(R) & = & N_V[V_{\rm IS}(R)- V_{\rm IV}(R)]
 -i[W_vf(R)-4a_wW_s\frac{df(R)}{dR}], \nonumber \\
\ {\rm where}\ f(R) & = & 1/\{1+\exp[(R-R_w)/a_w\}. \label{g6}
\end{eqnarray}
The normalization factor $N_V$ of the real folded potential as well as
strengths of the volume $W_v$ and surface $W_s$ parts of the
absorptive WS potential are adjusted to fit the elastic scattering data
at each energy. The WS radius $R_W$ and diffuseness $a_W$ were fixed as
taken from the CH89 global systematics \cite{Va91}. The OM descriptions 
of the elastic proton scattering from $^{48}$Ca, $^{90}$Zr, $^{120}$Sn, and 
$^{208}$Pb targets at 35 and 40 MeV given by the two choices of the proton OP 
are shown in Fig.~\ref{f4}, and the corresponding OP parameters are presented
in Tables~\ref{t1} and \ref{t2}. The best OM fit to the elastic data required
the strength of the complex folded potential to be renormalized by 
$N_{\rm V}\approx 0.8\sim 0.90$ and $N_{\rm W}\approx 0.5\sim 0.6$. An 
absorption overestimated by the folding model has been observed earlier 
\cite{Kho07} and it is due mainly to a strong volume absorption predicted 
by the JLM potential. In general, after the adjustment of the OP 
parameters, both the folded (\ref{g5}) and hybrid (\ref{g6}) optical 
potentials give equally good OM descriptions of the data. 
The calculated total reaction cross sections also 
agree nicely with the experimental data, and that confirms the reliability 
of the proton OP used in our CC equations for the \cx scattering. 
The coupling between the elastic and \cx scattering channels turned out 
to be quite weak and all the OP parameters in the CC calculation have been 
kept as fixed by the OM analysis of elastic proton scattering.   
\begin{table}
\caption{Best-fit renormalization factors $N_{V}$ and $N_{W}$ of the complex 
folded proton OP of the entrance channel (\ref{g5}). The calculated proton 
total reaction cross section $\sigma_{\rm R}$ is compared with the data
$\sigma_{\rm R}^{\rm exp}$ taken from Ref.~\cite{Ca96}. $N_{\rm V1(W1)}$ are
the renormalization factors of the folded \cx FF (\ref{g7}) deduced from the
best CC fit to the \pn data.} \label{t1}
\begin{tabular}{cccccccc}\hline
Target & $E_p$ (MeV) & $N_{V}$ & $N_{W}$ & $\sigma_R$ (mb) & $\sigma_R^{\rm exp}$ (mb)
 & $N_{V1}$ & $N_{W1}$ \\ \hline
$^{48}$Ca  & 35  & $0.817$ & $0.479$ & $959$  & $971\pm32 $ & $1.352$ & $1.00$ \\
           & 45  & $0.840$ & $0.515$ & $875$  & $908\pm34 $ & $1.430$ & $1.00$ \\
$^{90}$Zr  & 35  & $0.898$ & $0.551$ & $1357$ & $1316\pm65\ ^a$ & $1.516$ & $1.00$ \\
           & 45  & $0.784$ & $0.561$ & $1246$ & $1214\pm59\ ^b$ & $1.557$ & $1.00$ \\
$^{120}$Sn & 35  & $0.872$ & $0.503$ & $1583$ & $1668\pm59$ & $1.377$ & $1.00$ \\
           & 45  & $0.859$ & $0.521$ & $1508$ & $1545\pm38$ & $1.539$ & $1.00$ \\
$^{208}$Pb & 35  & $0.858$ & $0.494$ & $1934$ & $1974\pm38$ & $1.479$ & $1.00$ \\
           & 45  & $0.864$ & $0.538$ & $1988$ & $1979\pm41$ & $1.464$ & $1.00$ \\ \hline
\end{tabular}\\
$^a$ Total reaction cross section measured at $E=40$ MeV;
$^b$ at $E=49.5$ MeV. \\
\end{table}
\begin{table}
\caption{The same as Table~\ref{t1} but for the hybrid OP
(\ref{g6}), with the diffuseness of the WS potential $a_w=0.69$ fm for
all cases as given by the CH89 global OP \cite{Va91}.} \label{t2}
\begin{tabular}{cccccccccc} \hline
Target & $E_p$ & $N_V$ & $W_v$ & $W_s$ & $R_w$ & $\sigma_R$ 
 & $\sigma_R^{\rm exp}$ & $N_{V1}$ & $N_{W1}$ \\
   &(MeV)& & (MeV) & (MeV) & (fm) & (mb) & (mb) & & \\ \hline
$^{48}$Ca & 35 & 0.839 & 2.839 & 4.601 & 4.414 & 979 & $971\pm32$ & 1.370 & 1.00 \\
   & 45 & 0.847 & 5.107 & 3.265 & 4.414 & 908 & $908\pm34$ & 1.578 & 1.00 \\
$^{90}$Zr & 35 & 0.877 & 2.479 & 7.060 & 5.540 & 1419 & $1316\pm65$& 1.537 & 1.00 \\
   & 45 & 0.877 & 3.630 & 6.261 & 5.540 & 1400 & $1214\pm59$& 1.558 & 1.00 \\
$^{120}$Sn & 35 & 0.868 & 2.305 & 7.792 & 6.140 & 1639 & $1668\pm59$ & 1.340 & 
1.00 \\ 
   & 45 & 0.875 & 2.727 & 5.530 & 6.140 & 1545 & $1545\pm38$ & 1.357 & 1.00 \\
$^{208}$Pb & 35 & 0.868 & 4.982 & 5.473 & 7.460 & 1957 & $1979\pm41$ & 1.582 & 
1.00 \\
  & 45 & 0.873 & 6.625 & 3.338 & 7.460 & 2011 & $1974\pm38$ & 1.428 & 1.00 \\ \hline 
\end{tabular}
\end{table}
The complex OP of the exit channel is constructed using relation (\ref{e6b}), 
with the IS and IV parts of the neutron OP determined at the energy $E=E_p-Q$ as 
discussed above. Because the OP between neutron and nucleus in the IAS is 
unknown, we simply adjusted the renormalization factors $N_{V(W)}$ of the 
folded OP of the exit channel to the best CC description of the elastic and 
\pn scattering data. In all cases under study, the best-fit $N_{V(W)}$ factors
turned out to be $10\sim 15\%$ smaller than those obtained for the folded OP 
of the entrance channel shown in Table~\ref{t1}.  

\begin{figure}[bht] \vspace*{-1cm}\hspace{-1cm}
\includegraphics[width=1.2\textwidth]{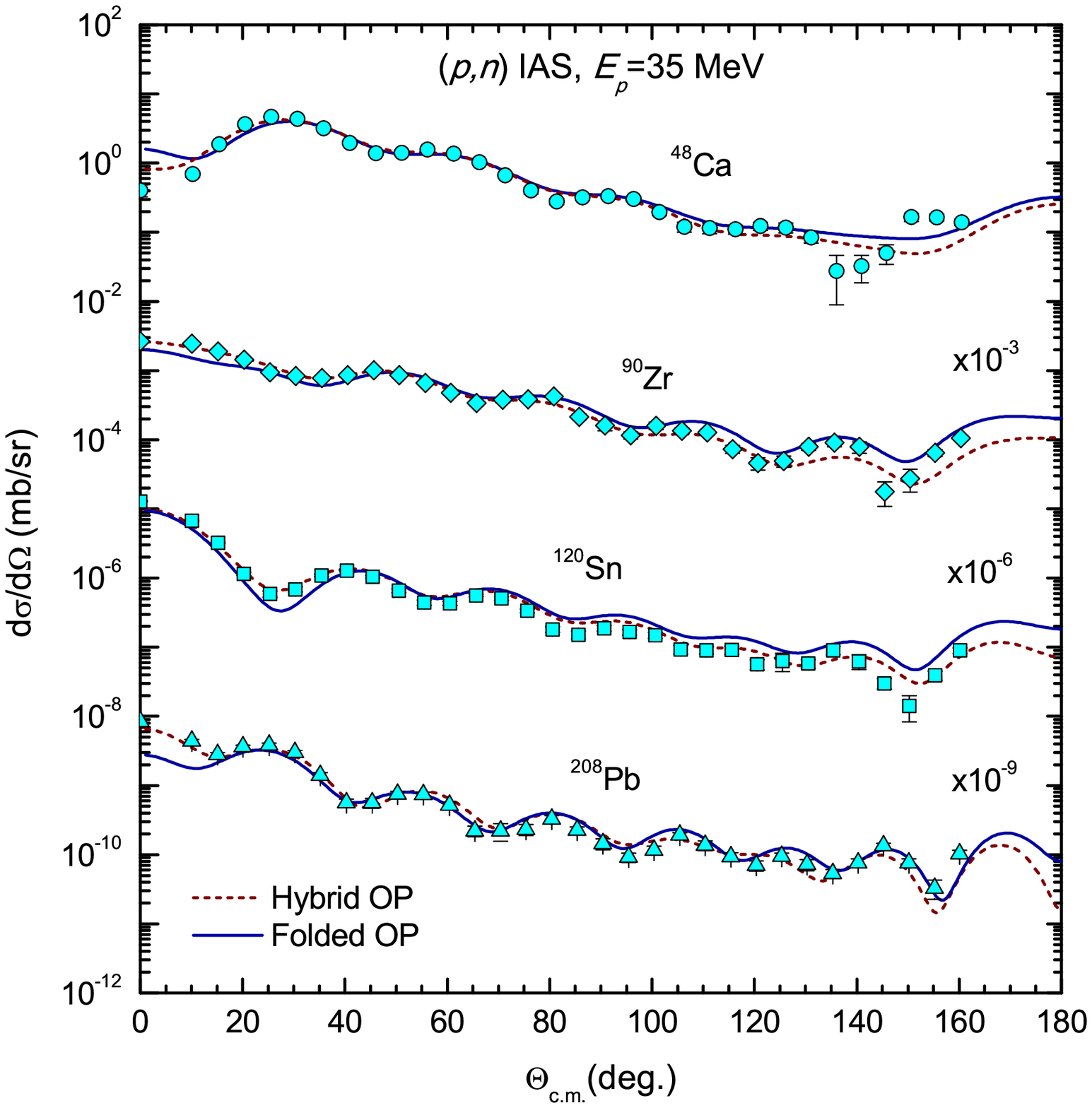}\vspace*{-1cm}
 \caption{CC description of the \cx \pn scattering to the IAS of  
$^{48}$Ca, $^{90}$Zr, $^{120}$Sn, and $^{208}$Pb targets at 
$E_p=35$ MeV given by the \cx form factor (\ref{g7}) used with 
the complex folded OP or hybrid OP. The data were taken from 
Ref.~\cite{Doe75}.}  
 \label{f5}
\end{figure}
\begin{figure}[bht] \vspace*{-1cm}\hspace{-1cm}
\includegraphics[width=1.2\textwidth]{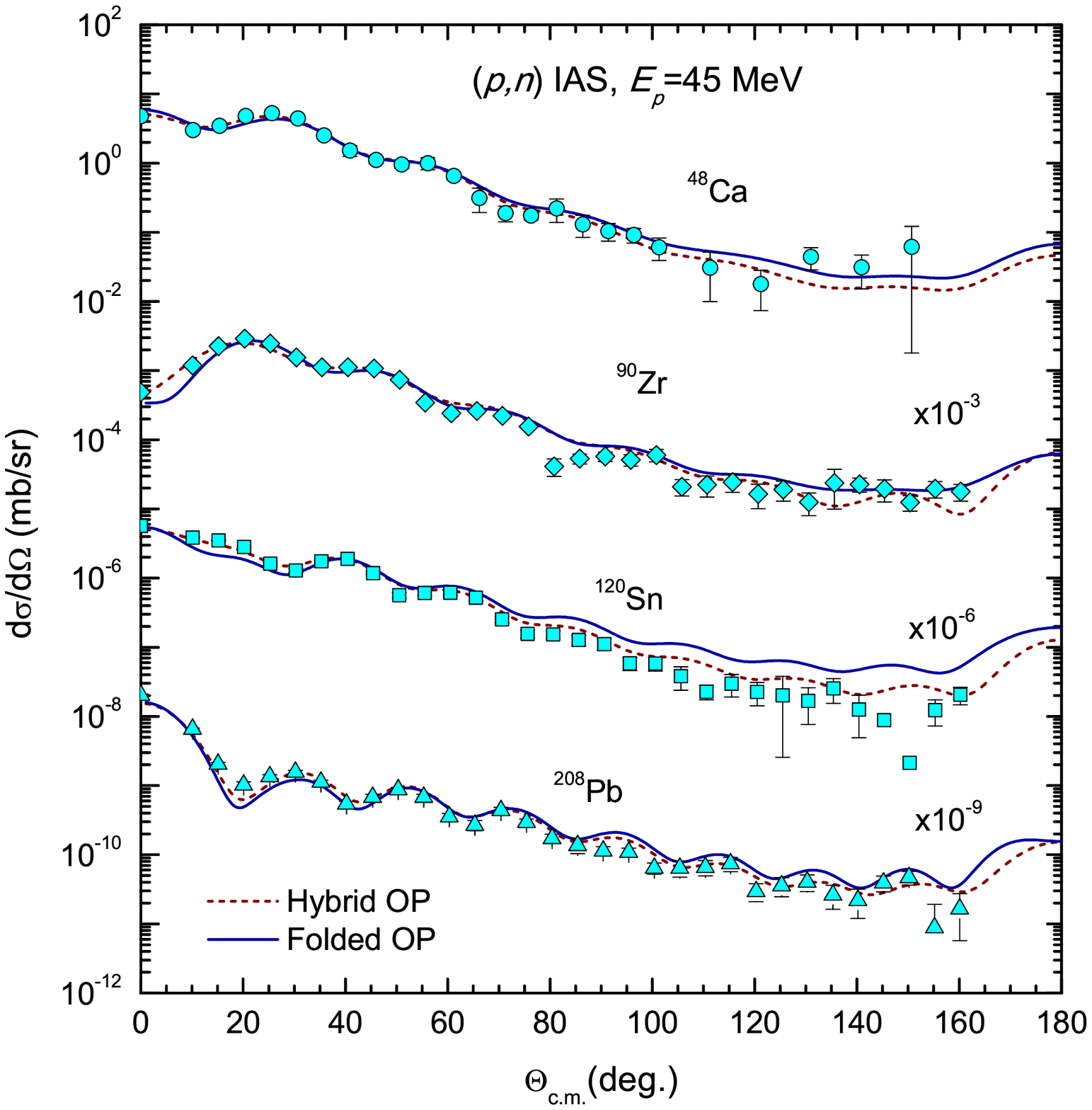}\vspace*{-1cm}
 \caption{The same as Fig.~\ref{f5} but for $E_p=45$ MeV.}  
 \label{f6}
\end{figure}

As already noted, the isovector strength of the proton OP is just a few 
percent of the total OP and it is difficult to adjust the strength 
of $U_{\rm IV}$ separately in an OM fit to the elastic scattering data.
However, the IV strength of the OP can be very well fine tuned against the 
\pn data in the DWBA or CC analysis of the \cx scattering because the FF 
of the \cx scattering to the IAS is entirely determined by $U_{\rm IV}$. 
Thus, the following complex \cx FF is used in our CC analysis 
\begin{equation}
 F_{\rm cx}(R)=\sqrt{\frac{2}{T_z}}U_{\rm IV}(R)=\sqrt{\frac{2}{T_z}}
 [N_{V1}V_{\rm IV}(R)+ iN_{W1}W_{\rm IV}(R)], \label{g7}
\end{equation}
where $V_{\rm IV}(R)$ and $W_{\rm IV}(R)$ are the folded IV potential 
obtained with the CDM3Y6 interaction determined at $E=E_p-\Delta E_C-Q/2$.
The CC results for the \cx \pn scattering to the IAS of  $^{48}$Ca, $^{90}$Zr, 
$^{120}$Sn, and $^{208}$Pb targets at $E_p=35$ and 45 MeV given by the 
\cx form factor (\ref{g7}) are shown in Figs.~\ref{f5} and \ref{f6}, and the 
corresponding best-fit renormalization factors $N_{V1}$ and $N_{W1}$ are 
given in Tables~\ref{t1} and \ref{t2}.
One can see that both choices of the proton OP give about the same good CC
fit to the \pn data, with a slight improvement by the more elaborate hybrid OP.
The CC results obtained with both choices of the proton OP clearly show
that the best fit to the \pn data required the complex form factor (\ref{g7}) 
to be renormalized by  $N_{V1}\sim 1.3-1.5$ and $N_{W1}\sim 1.0$. This  
indicates that the empirical IV strength of the CDM3Y6 interaction 
(\ref{g3})-(\ref{g4}) should be about $30\sim 40\%$ stronger than that 
adjusted to the JLM nucleon OP in nuclear matter limit \cite{Je77}.
It must be  noted that the Coulomb correction to the proton incident 
energy $\Delta E_C$ has not been taken into account in our earlier folding model 
analysis \cite{Kho07} of these same \pn data, where the fully folded OP could  
deliver a fair description of the \cx scattering data only with $N_{V1}\sim 2.0$ 
(much larger than that obtained in the present work). In the present work,
a good CC description of the \pn data has been reached with the fully folded 
OP, using about the same scaling factors $N_{V1(W1)}$ as those obtained with 
the hybrid OP. These results show that the Coulomb correction is 
important not only in the OM analysis of the nucleon scattering but also 
in a Lane-consistent CC description of the elastic and \cx scattering 
to the IAS \cite{deVito}. Our CC results give a good description not only to the
measured angular distribution of the \cx \pn scattering but also to the total 
\pn cross section. For example, the CC calculation using the hybrid OP and 
renormalized complex folded FF gives the total \pn cross section
$\sigma_{pn}=4.8$ and 4.1 mb for the \pnZr reaction at 35 and 45 MeV,
respectively, which agree nicely with the data ($\sigma_{pn}^{\rm exp}\approx
4.8\pm 0.5$ and $4.4\pm 0.5$ mb at 35 MeV and 45 MeV, respectively)
\cite{Doe75}.

It is complimentary to note that the effective JLM interaction of Gaussian 
type has been used earlier by Pakou {\sl et al.} \cite{Pak01} and Bauge 
{\it et al.} \cite{Bau01} to study the same \pn reactions. The JLM folding model 
analysis of the proton, neutron elastic scattering and \cx \pn reaction done 
in Ref.~\cite{Pak01} has also shown that the IV strength of the JLM interaction 
is too weak and a strong overall renormalization of the folded \cx FF by 
$N_{\rm V1}=N_{\rm W1}\approx 2-2.5$ was needed to account for the 
\pn data. In a more elaborate treatment of the \cx transition to the IAS 
\cite{Bau01} the isospin coupling factor in Eq.~(\ref{e3b}) has been assumed 
density dependent, $\sqrt{2T_z/A}=\sqrt{[\rho_n(r)-\rho_p(r)]/\rho(r)}$. 
The JLM nucleon OP obtained with such a density-dependent isospin coupling 
has been thoroughly tested in the OM analysis of the elastic and \pn
scattering data measured over a wide range of energies and target masses 
\cite{Bau01}, and the best-fit renormalization factors $N_{\rm V1}\approx
1.5-1.6$ and $N_{\rm W1}\approx 1.3-1.4$ were found for the \cx folded 
FF in the energy range of $30-40$ MeV, which are close to our results. 
Thus, all the results show consistently that the IV strength of the 
JLM interaction is much too weak to account for the measured \pn data. 
Because the IV term of the JLM nucleon OP has been obtained as the 
first-order expansion of the mass operator of symmetric NM perturbed by a 
neutron excess \cite{JLM77}, a weakness of the resulting JLM nucleon OP in 
asymmetric NM could well be expected. We conclude here that the \cx scattering 
to the IAS is indeed a very helpful reaction to probe the isospin dependence 
of an effective NN interaction. It would be of interest, therefore, to apply 
similar folding model analysis to test the isospin dependence of the nucleon 
OP given by the advanced BHF calculation of asymmetric NM \cite{Zuo06}.

\subsection{Implication for the symmetry energy of nuclear matter}
\label{sse}
As discussed above, the knowledge about the isospin dependence of the in-medium
NN interaction is vital for the construction of the EOS of asymmetric NM, 
the key input for the studies of neutron star \cite{Bet90,Lat04,Ste05}. 
We show here that the results of the folding model analysis of the \cx 
\pn scattering are quite helpful for the determination 
of the NM symmetry energy. For this purpose, the real IV density 
dependence (\ref{g4}) of the CDM3Y6 interaction at low energies 
(keV region) has been carefully parametrized to reproduce in the HF 
approximation the IV part of the nucleon OP in NM limit given by the 
microscopic BHF calculation of the JLM group \cite{Lej80}. Such a version 
of the density- and isospin dependent CDM3Y6 interaction is then used in the 
HF calculation \cite{Kho96,Than09} of the total energy density 
$\mathcal{E}$ of the asymmetric NM  
\begin{equation}
\mathcal{E}=\mathcal{E}_{\rm kin}+{\frac{1}{ 2}}\sum_{k \sigma \tau}
\sum_{k'\sigma '\tau '} [\langle{\bm k}\sigma \tau, {\bm k}' \sigma' \tau'
 |v_{\rm D}|{\bm k}\sigma\tau, {\bm k}' \sigma' \tau' \rangle
+ \langle{\bm k}\sigma \tau, {\bm k}'\sigma' \tau' |v_{\rm EX}|
{\bm k}'\sigma \tau, {\bm k}\sigma' \tau' \rangle],
 \label{h1} 
\end{equation}
where $|{\bm k}\sigma \tau\rangle$ are the ordinary plane waves. Dividing 
$\mathcal{E}$ over the total NM density $\rho$, we obtain the total 
NM energy per particle $E$
\begin{equation}
\frac{\mathcal{E}}{\rho}=E(\rho,\delta)=
 E(\rho,\delta=0)+ S(\rho)\delta^2+O(\delta^4)+...,\ 
 \delta=\frac{\rho_n-\rho_p}{\rho}.  \label{h2}
\end{equation}
The contribution of $O(\delta^4)$ and higher-order terms in 
the neutron-proton asymmetry $\delta$ has been proven to be small 
\cite{Kho96,Zuo99} and is neglected in the \emph{parabolic} approximation, 
where the NM symmetry energy $S(\rho)$ simply equals the energy required 
per particle to change the symmetric NM into the pure neutron matter. 
The value of $S(\rho_0)$ at the saturation density ($\rho_0\approx 0.17$\ 
fm$^{-3}$) has been predicted by different many-body calculations to be 
around 30-31 MeV \cite{Kho96,Zuo99,Bra85,Pea00}.

\begin{figure}[bht] \vspace*{-1cm}
\includegraphics[width=\textwidth]{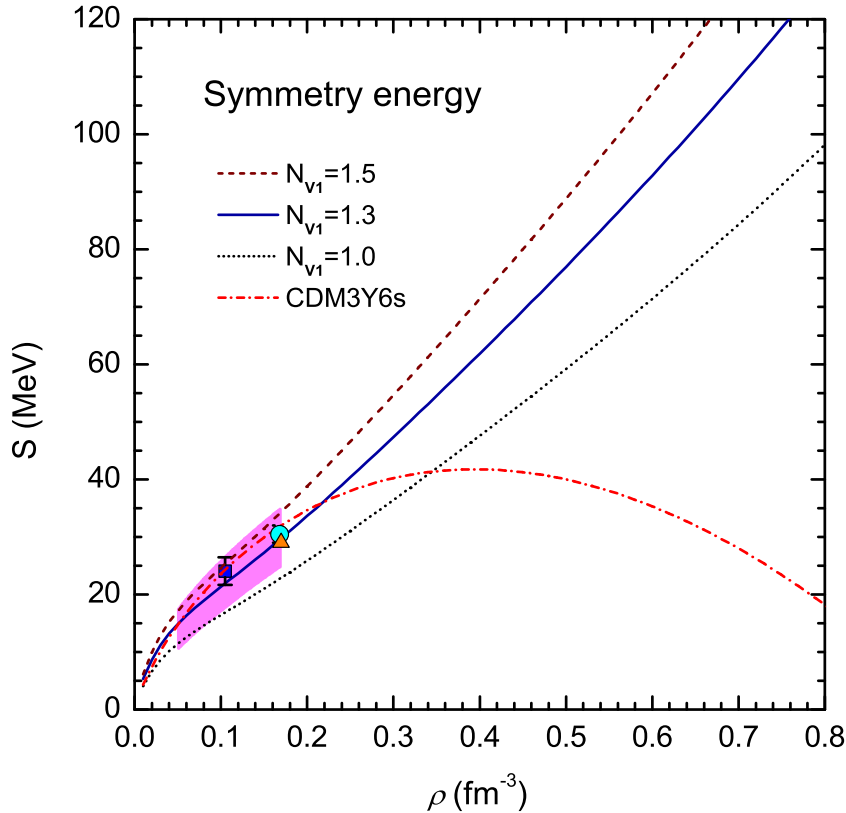}\vspace*{-2cm}
\caption{HF results for the nuclear symmetry energy $S(\rho)$ given by the  
CDM3Y6 interaction, with the IV density dependence renormalized by different 
factors $N_{V1}$. The shaded (magenta) region marks the empirical boundaries 
implied by the isospin diffusion data and double ratio of neutron and proton 
spectra of HI collisions \cite{Tsa09,Ono03}. The circle is the empirical value 
predicted by nuclear many-body calculations \cite{Kho96,Zuo99,Bra85,Pea00} that 
is about the same as that value deduced from the folding model analysis
of the \cx \pn data \cite{Kho05,Kho07}. The square and triangle are the 
constraints deduced from the consistent structure studies of the giant dipole 
resonance \cite{Tri08} and neutron skin \cite{Fur02}, respectively. CDM3Y6s
is a ``soft" version of the CDM3Y6 interaction, with about the same density 
dependence assumed for both the IS and IV terms \cite{Loan11}.} \label{f7}
\end{figure}
Our HF results for the nuclear symmetry energy $S(\rho)$ given by different
IV strengths (\ref{g4}) of the real CDM3Y6 interaction  are shown in 
Fig.~\ref{f7}. One can see that $S(\rho_0)$ is approaching the empirical 
value of around $30-31$ MeV only if Re~$F^{\rm V}_{\rm IV}(\rho)$ is 
renormalized by $N_{V1}\approx 1.3-1.5$, in a good agreement with the 
renormalization factors given by the folding model analysis of the \cx \pn 
data. The use of the \emph{unrenormalized} IV density dependence based on 
the JLM results clearly underestimates $S(\rho_0)$ compared to the 
empirical values. The weakness of the JLM isovector strength \cite{Je77} is, 
thus, also confirmed in our HF results for asymmetric NM. The IV density 
dependence of the CDM3Y6 interaction renormalized by $N_{V1}\approx 1.3-1.5$ 
also give $S(\rho)$ values lying within the empirical boundaries implied 
by the HI fragmentation data \cite{Tsa09,Ono03,She07,Sh07} and the nuclear 
structure studies of the giant dipole resonance \cite{Tri08} and neutron skin 
\cite{Fur02}, at the NM densities up to $\rho_0$.  

\begin{figure}[bht] \vspace*{-1cm}\hspace{0cm}
\includegraphics[width=\textwidth]{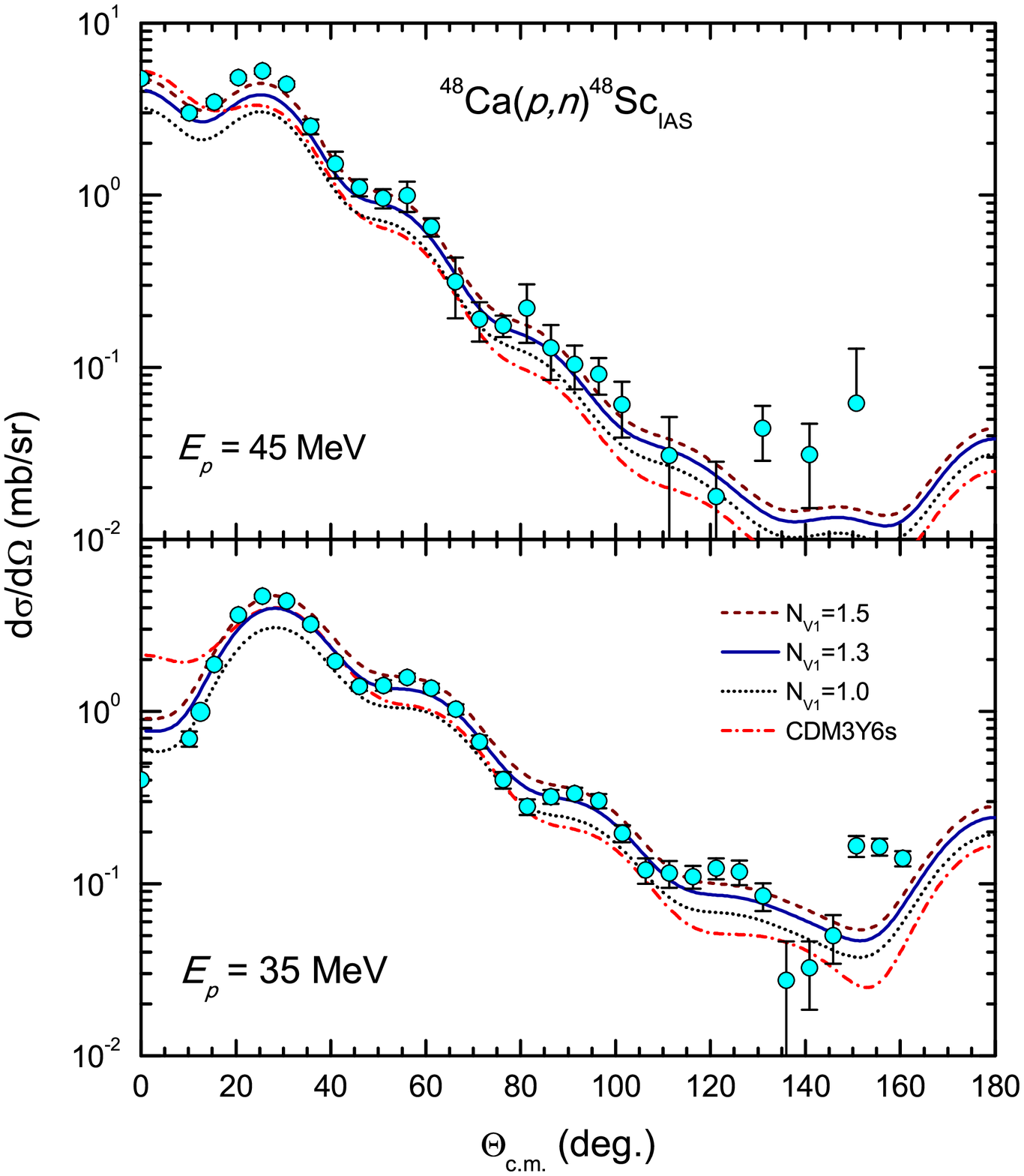}\vspace*{-1.5cm}
 \caption{CC description of the \cx \pn scattering to the IAS of  
$^{48}$Ca target at $E_p=35$ and 45 MeV given by the hybrid OP and 
\cx form factor (\ref{g7}) obtained with the different inputs for the IV 
density dependence of the CDM3Y6 interaction, similar to those discussed 
in Fig.~\ref{f7}. The data were taken from Ref.~\cite{Doe75}.}  
\label{f8}
\end{figure}
Although the folding model analysis of the \pn scattering to the IAS
has put a constraint on the nuclear symmetry energy $S(\rho)$ at 
$\rho\lesssim \rho_0$ as shown in Fig.~\ref{f7}, its behavior at 
higher NM densities remains uncertain due to a simple reason that 
the total nuclear density of the \pA system never exceeds $\rho_0$, 
so that the \pn data are sensitive mainly to the low-density tail 
of the isovector interaction. 
To explore this effect in more details, we have assumed for the IV 
density dependence $F_{\rm IV}(\rho)$ of the CDM3Y6 interaction the 
same density-dependent functional as that of the IS density 
dependence but scaled by a factor of 1.1 deduced from our earlier 
folding model analysis of the \h6pn reaction \cite{Kho05}, i.e.,
$F_{\rm IV}(\rho)=1.1F_{\rm IS}(\rho)$. Such an ansatz for the 
IV density dependence of the CDM3Y6 interaction, dubbed as the 
CDM3Y6s interaction \cite{Loan11}, gives nearly the same description 
of the NM symmetry energy $S(\rho)$ at $\rho\lesssim \rho_0$ as the newly 
parametrized $F_{\rm IV}(\rho)$,
with a slight difference in the slope of the corresponding $S(\rho)$
curves (compare dash-dotted and solid curves in Fig.~\ref{f7}). These 
two sets of the IV density dependence lead, however, to very different 
behaviors of $S(\rho)$ at high NM densities. The symmetry energy obtained with 
$F_{\rm IV}(\rho)$ based on the JLM results increases monotonically with the
increasing NM density. Such a behavior has been widely discussed in the 
literature as the \emph{stiff} density dependence of the NM symmetry energy. 
The behavior of $S(\rho)$ obtained with the CDM3Y6s interaction is 
referred to as the \emph{soft} density dependence \cite{Than09,Loan11,Bar05}. 

\begin{figure}[bht] \hspace*{-0.5cm}
\includegraphics[width=1.1\textwidth]{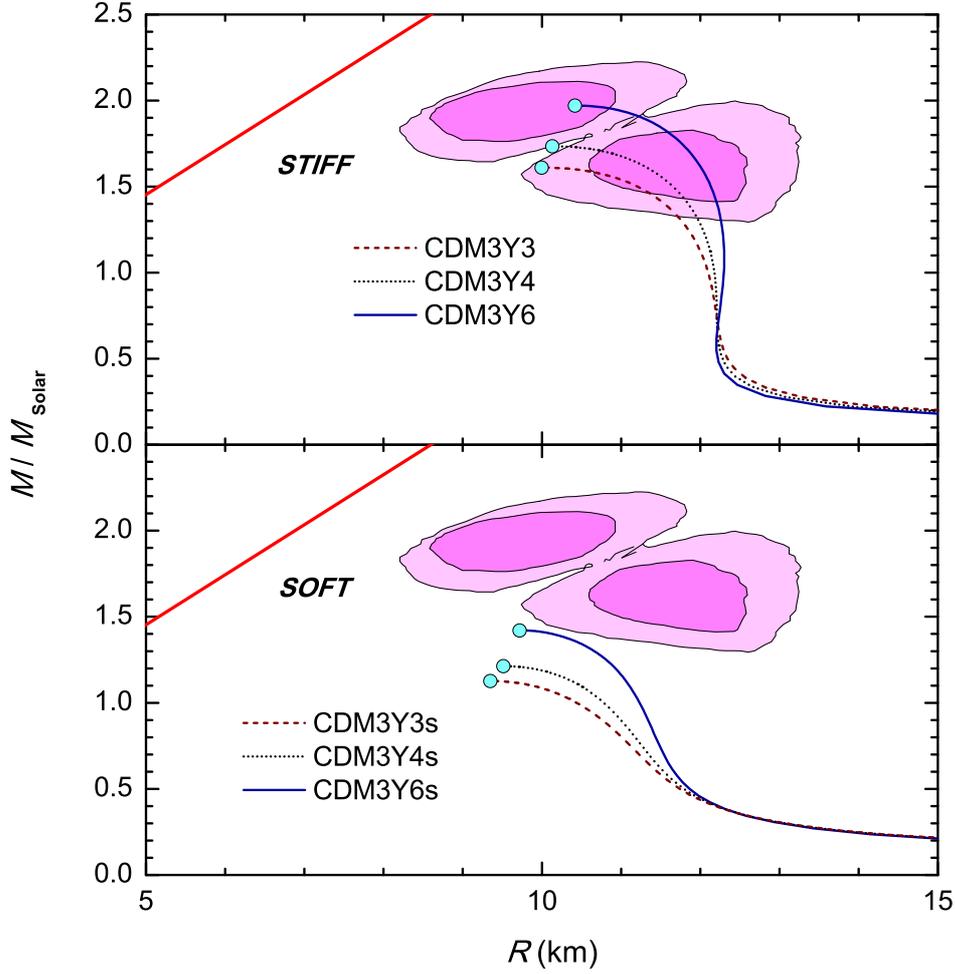}\vspace*{-0.5cm}
 \caption{The gravitational mass of neutron star versus its radius obtained 
with the EOS's given by the stiff-type (upper panel) and soft-type (lower panel) 
CDM3Yn interactions, in comparison with the empirical data (shaded contours) 
deduced by Steiner {\it et al.} \cite{Ste10} from the observation of the X-ray 
burster 4U 1608-52. The circles are values calculated at the maximum central 
densities. The thick solid (red) line is the limit allowed by the 
General Relativity \cite{Glen2}.} \label{f9}
\end{figure}
To see if the \pn data under study are sensitive to the slope 
of the NM symmetry energy in the low-density region, we have done the 
CC analysis of the \cx \pn scattering on $^{48}$Ca target at $E_p=35$ 
and 45 MeV using the form factors (\ref{g7}) given by different choices 
of the IV density dependence of the CDM3Y6 interaction, similar to those 
used in the HF calculation of the NM symmetry energy shown in Fig.~\ref{f7}. 
In particular, the (soft) CDM3Y6s version of the interaction has 
Re$F_{\rm IV}(\rho)=1.1\times$~Re$F_{\rm IS}(\rho)$ and 
Im$F_{\rm IV}(\rho)=$~Im$F_{\rm IS}(\rho)$. One can see in Fig.~\ref{f8}
that the use of the soft IV density dependence leads to a much poorer 
description of the \pn data at large angles. Although nearly the same 
description of the data at forward angles is given by both stiff and soft 
choices of the IV density dependence, the 35 MeV data points approaching 
the zero angle cannot be properly accounted for by the FF obtained with 
the soft CDM3Y6s interaction. This result shows that a difference in the 
slope of the NM symmetry energy at low NM densities shown in Fig.~\ref{f7} 
can be traced in the calculated \pn cross section, and the best CC
fit to the \pn data prefers the stiff IV density dependence of the 
CDM3Y6 interaction. We note that some HI collision data also 
were found to prefer the stiff density dependence of the NM symmetry 
energy \cite{Bar05}. 

A similar approach has been used in our recent study of the EOS of
neutron star matter \cite{Loan11} with the density dependent CDM3Y3, 
CDM3Y4 and CDM3Y6 interactions. The IS density dependence 
$F_{\rm IS}(\rho)$ of these three versions of the density dependent 
M3Y interaction has been parametrized \cite{Kho07r,Kho97} to reproduce 
the empirical saturation properties of symmetric NM in the HF 
approximation, with the incompressibility $K=217$, 228 and 252 MeV, 
respectively. Their IV density dependence $F_{\rm IV}(\rho)$ has been 
determined also in two (stiff and soft) scenarios, in exactly the same 
way as discussed above. The stiff and soft versions of the density 
dependent M3Y interaction give about the same behavior of the 
NM symmetry energy as shown in Fig.~\ref{f7} for the CDM3Y6 and 
CDM3Y6s interactions. The EOS of the uniform core of neutron 
star of the $npe\mu$ composition in the $\beta$-equilibrium at zero 
temperature has been calculated using the stiff and soft versions of 
the density dependent NN interaction \cite{Loan11}. The obtained EOS's 
were then used as input of the Tolman-Oppenheimer-Volkov equations 
to describe basic properties of neutron star, like the gravitational mass, 
radius and moment of inertia. The most obvious effect caused by changing 
slope of the NM symmetry energy from \emph{stiff} to \emph{soft} is the 
reduction of the maximum gravitational mass $M$ and radius $R$ as 
illustrated in Fig.~\ref{f9}. Namely, the $M$ value is changing from 
$1.6\sim 2\ M_\odot$ to a significantly lower range of $1.1\sim 1.4\ M_\odot$, 
with a much worse description of the empirical mass-radius data 
\cite{Ste10}. Together with the present results of the folding model
analysis of the \cx scattering to the IAS, it is highly plausible that one
can rule out the EOS with a \emph{soft} behavior of the symmetry energy
in the theoretical modeling of neutron star.

\subsection{Folding model analysis of the \he3t scattering to the IAS}
As mentioned in Sec.~\ref{sec1}, a folding model study the $(^3$He,$t$)
scattering to the IAS might allow us to test the high density part 
of the isovector density dependence (\ref{g4}), due to a higher 
overlap nuclear density reached during the collision and, eventually, 
to conclude on the slope of the NM symmetry energy $S(\rho)$ 
at higher densities.

Given the same spin and isospin of proton and $^3$He, very similar 
structures of the initial and final states have been observed in the 
\cx \pn and \he3t reactions. In particular, the \cx scattering to the 
IAS induced by both proton and $^3$He projectiles can be considered in
the same scenario of the isospin-flip elastic scattering 
\cite{Dri62,Sat64,Sat69} caused by the isovector term (Lane potential) 
of the optical potential (\ref{e1}). Therefore, the determination 
of the Lane potential $U_1$ has been attempted already in the \he3t 
experiments in the early 70's \cite{Hin72,Hay74}. However, due to 
the complexity of the \he3t reaction that is caused by a composite 
projectile, the isospin dependence of the \heA OP is much less known 
compared to that of the nucleon OP. Unlike the \nA case, the isovector 
term of the real \heA OP could not even be established in the recent 
global OP for $^3$He and triton \cite{Pang09}. It is difficult, 
therefore, to investigate the Lane consistency of the phenomenological
\heA OP based on the DWBA or CC analysis of the elastic and \cx 
scattering as has been done in Ref.~\cite{Kho07} with the 
phenomenological global nucleon OP. On the microscopic level, the 
measured data of the \he3t scattering to the IAS have been analyzed mainly
in the DWBA with the \cx form factor given by a single-folding calculation
using an effective $^3$He-nucleon interaction and nuclear density
of the target \cite{Werf1,Werf2}. Therefore, it is of high interest to carry
out a consistent folding model study of both the \pn and \he3t scattering to the 
IAS using the same density- and isospin dependent NN interaction. 

\begin{figure}[bht] \vspace*{0cm}\hspace{-1cm}
\includegraphics[width=1.2\textwidth]{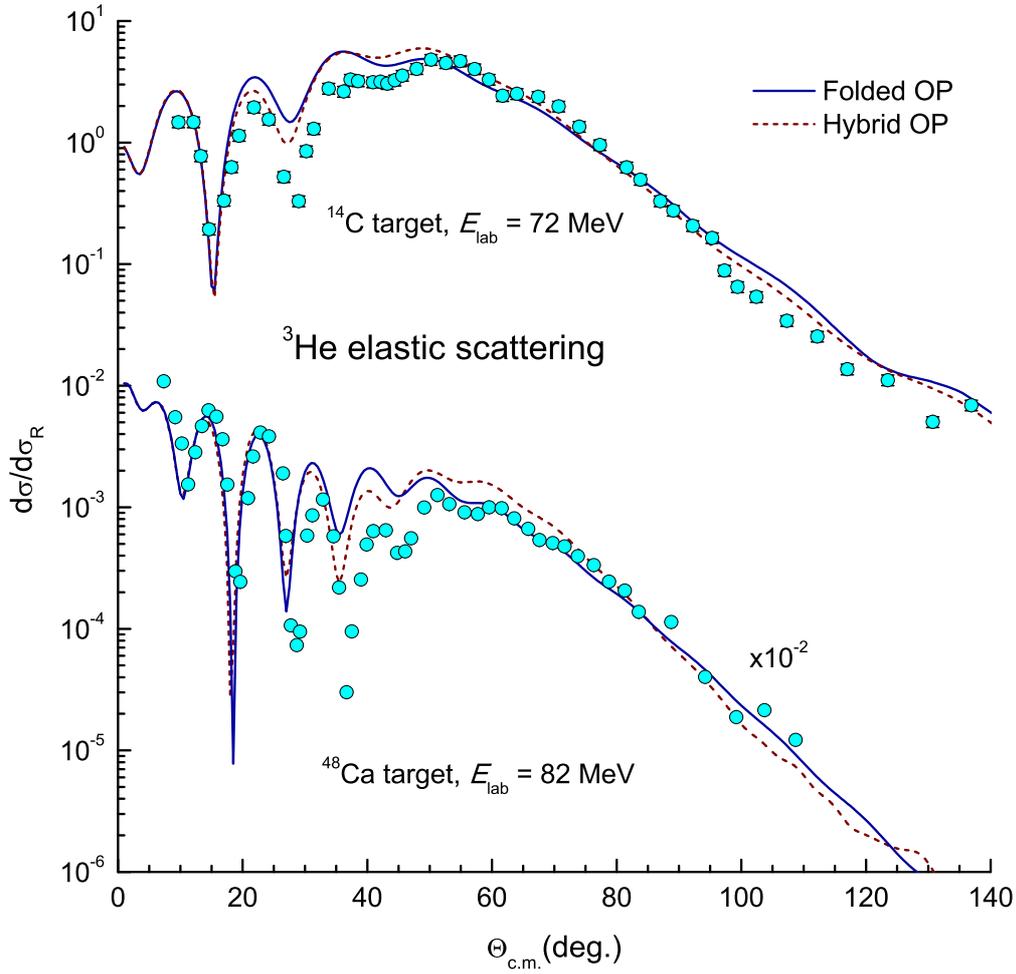}\vspace*{0.5cm}
\caption{CC description of the elastic $^3$He+$^{14}$C and 
$^3$He+$^{48}$Ca scattering data at $E_{\rm lab}=72$ MeV \cite{Alla} and 
82 MeV \cite{Tanabe}, respectively, given by the complex folded and 
hybrid optical potentials (\ref{g5})-(\ref{g6}).} 
 \label{f10}
\end{figure}

\begin{figure}[bht] \vspace*{0cm}\hspace{-1cm}
\includegraphics[width=1.2\textwidth]{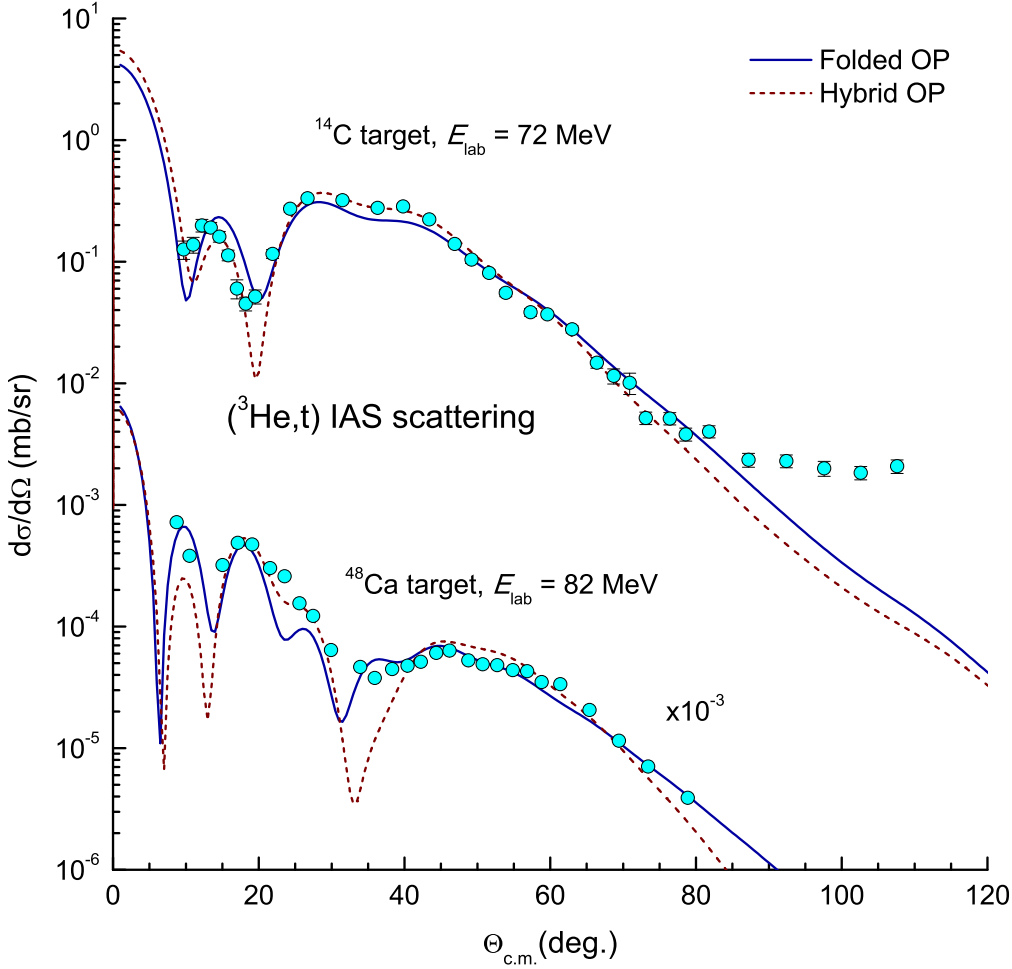}\vspace*{0.5cm}
\caption{CC description of the \cx \xc and \xca scattering to the IAS 
of the target at $E_{\rm lab}=72$ and 82 MeV, respectively, given 
by the folded form factor (\ref{g7}) used with fully folded and hybrid 
optical potentials (\ref{g5})-(\ref{g6}). The data were taken from 
Refs.~\cite{Alla,Tanabe}.} 
 \label{f11}
\end{figure}

In the present work we have considered two representative cases
of the \he3t scattering to the IAS: \xc and \xca data measured at 
$E_{\rm lab}=72$ MeV \cite{Alla} and 82 MeV \cite{Tanabe}, respectively.
Both data sets were measured together with the elastic $^3$He scattering,
over a wide angular range \cite{Kho07r,Alla}.  
The $^3$He incident energies of 72 and 82 MeV (or 24 and 27.3 MeV/nucleon)
are in the range of the ``rainbow" energy \cite{Kho07r} and one can see
in Figs.~\ref{f10} and \ref{f11} that both the elastic and \cx data show 
clearly the shoulder-like bump characteristic of the nuclear rainbow
\cite{Kho07r} at large scattering angles. Therefore, the elastic and 
\he3t scattering data under study should be quite sensitive to the 
strength and shape of the real OP and, hence, could serve as a probe
of the folding model used to predict the OP. The two considered energies 
are also not too low so that the \he3t scattering to the IAS can be 
treated as a direct one-step process \cite{Alla,Tanabe}. 

Like the \pn scattering, the isospin impurity of the \heA OP
needs to be taken into account by a correction $\Delta E_C$ to the 
incident $^3$He energy. The Coulomb correction to the global OP for 
$^3$He and triton used by Pang {\it et al.} \cite{Pang09} is
based on the CH89 systematics \cite{Va91}, i.e., 
$\Delta E_C=6Z_1Z_2e^2/(5R_C)$. Such an empirical estimate of the
Coulomb correction is also used in our folding model analysis. 
Thus, the folded \heA OP of the entrance channel (\ref{e6a}) is obtained 
with the complex CDM3Y6 interaction determined at the (Coulomb corrected) 
incident energy per nucleon $E=E_{\rm lab}/3-\Delta E_C/2$. The \cx form 
factor (\ref{e15}) of the \he3t scattering to the IAS is given by the 
isovector part of the folded \heA OP obtained with the CDM3Y6 
interaction determined at $E=E_{\rm lab}/3-\Delta E_C/2-Q/6$. The 
folded triton OP of the exit channel(\ref{e6b}) is obtained with the 
CDM3Y6 interaction determined at the energy per nucleon 
$E=E_{\rm lab}/3-\Delta E_C(t)-Q/3$, where $\Delta E_C (t)$ is the 
Coulomb correction determined for triton using the same CH89-based 
formula \cite{Pang09}. The complex folded OP and hybrid OP are used in 
the same notation as (\ref{g5})-(\ref{g6}), with parameters determined 
from the best CC fit to the elastic and \he3t scattering data as done 
in the CC analysis of the \pn scattering. The $^3$He and triton optical
potentials are added by the standard Coulomb and spin-orbital terms 
taken from the global OP by Pang {\it et al.} \cite{Pang09}. The OP 
parameters and renormalization factors of the \cx form factor (\ref{g7}) 
given by the best CC fit to the elastic and \he3t scattering data are 
given in Table~\ref{t3}, and the comparison of the calculated scattering 
cross sections with the data are shown in Figs.~\ref{f10} and \ref{f11}. 
\begin{table}
\caption{Parameters of the optical potentials (\ref{g5})-(\ref{g6}) 
and \cx form factor (\ref{g7}) for $^3$He+$^{14}$C and 
$^3$He+$^{48}$Ca systems at $E_{\rm lab}=72$ and 82 MeV, respectively, 
given by the best CC fit to the elastic and \he3t scattering data 
\cite{Alla,Tanabe}. Parameters $R_w$ and $a_w$ of the hybrid OP 
were kept unchanged as taken from the global OP by Pang {\it et al.} 
\cite{Pang09}.}  \label{t3}
\begin{tabular}{cccccccccc} \hline
Target & OP & $N_V$ & $W_v$ & $W_s$ & $R_w$ & $a_w$ & $\sigma_R$ 
 & $N_{V1}$ & $N_{W1}$ \\
   &  &  & (MeV) & (MeV) & (fm) & (fm) & (mb) & & \\ \hline
$^{14}$C & Folded$^*$ & 0.977 & 1.214 & - & - & - & 1009 & 1.300 & 0.850 \\
   & Hybrid & 1.054 & 5.786 & 10.175 & 3.027 & 0.84 & 1042 & 1.350 & 0.850 \\
$^{48}$Ca & Folded$^*$ & 0.889 & 1.433 & - & - & - & 1632 & 1.650 & 1.340 \\
   & Hybrid & 0.981 & 9.870 & 12.016 & 4.631 & 0.84 & 1713 & 1.420 & 1.280 \\
\hline 
\end{tabular}\\
$^*\ W_v$ is the best-fit factor $N_W$. \\
\end{table}
Like the folding analysis of the \pn data, we have treated 
the renormalization factors $N_{V(W)}$ of the folded triton OP of the 
exit channel as free parameters in our CC fit to the elastic and \he3t 
scattering data, and the best-fit $N_{V(W)}$ values also turned out 
to be $10\sim 15\%$ smaller than those obtained for the entrance 
channel. A similar effect of the strength reduction of the triton OP 
has also been found earlier in the DWBA analysis of the \he3t 
scattering to the IAS \cite{Werf1,Werf2}, using quite different 
model for the \cx FF.    

The double-folding approach (\ref{e13})-(\ref{e14}) using the 
density- and isospin dependent CDM3Y6 interaction gives a good 
prediction of the real \heA OP, with the best-fit renormalization 
factors $N_V\approx 0.9\sim 1$. The renormalization factors $N_W$ 
of the imaginary folded OP is around $1.2\sim 1.4$, and this 
is likely due to the lack of a realistic surface absorption (caused
by inelastic scattering to the low-lying excited states as well as
nucleon transfers) that cannot be predicted by the HF-type 
double-folded potential (\ref{e13})-(\ref{e14}). Nevertheless, 
the renormalized fully folded OP gives quite a good description of
the elastic scattering data. In the case of $^{48}$Ca target, the
fully folded OP gives a slightly better description of the elastic
data points in the rainbow region compared to that given by the 
hybrid OP (see lower panel of Fig.~\ref{f10}). The fact that the 
best-fit renormalization factors $N_V$ is quite close to unity shows 
that the real folded OP belongs indeed to the realistic potential
family for the elastic \heA scattering.       

\begin{figure}[bht] \vspace*{0cm}\hspace{-1cm}
\includegraphics[width=1.2\textwidth]{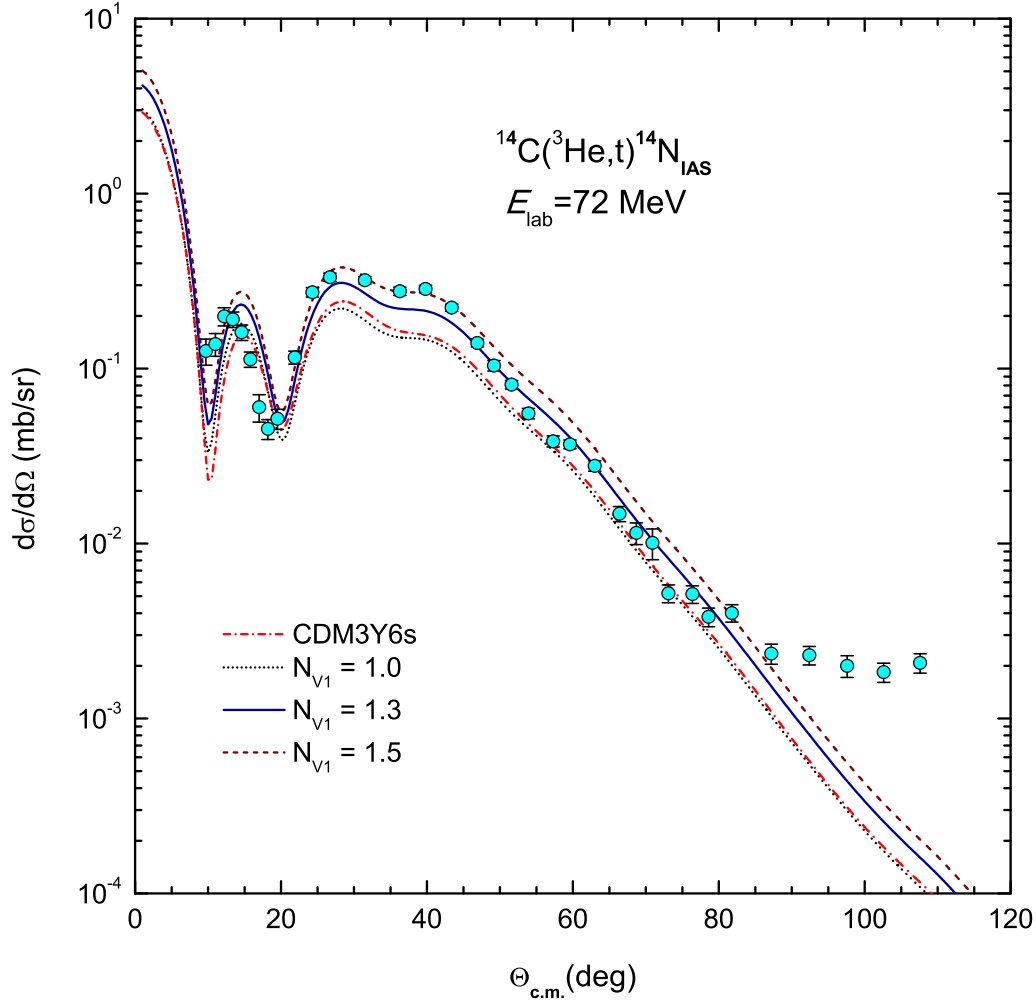}\vspace*{0cm}
\caption{CC description of the \cx \xc scattering data at 
$E_{\rm lab}=72$ MeV \cite{Alla} given by the folded OP (\ref{g5}) and 
\cx form factor (\ref{g7}) obtained with the different inputs for the IV 
density dependence of the CDM3Y6 interaction, similar to those discussed 
in Fig.~\ref{f7}.} 
 \label{f12}
\end{figure}

The CC results obtained with both choices of the $^3$He optical potential 
show that the best CC fit to the \cx \xc and \xca scattering data 
(see Figs.~\ref{f11} and \ref{f12}) needed a complex form factor (\ref{g7}) 
renormalized by $N_{V1}\approx 1.3\sim 1.6$ and $N_{W1}\approx 0.85\sim 1.3$. 
Such a behavior of the complex renormalization factor of the \cx FF is 
similar to that found above in the folding model analysis of the 
\pn scattering. This result confirms again that the empirical IV strength 
of the CDM3Y6 interaction (\ref{g3})-(\ref{g4}) should be $30\sim 40\%$ 
stronger than that adjusted to the JLM results for nuclear matter \cite{Je77}.

An important feature of the nuclear rainbow scattering is that the 
large-angle scattering data are quite sensitive to the \AA OP at small 
internuclear radii \cite{Kho07r}. In a folding model analysis, this means 
a sensitivity to the effective NN interaction at high nuclear densities. 
To see if the considered data of the \cx \he3t scattering to the IAS are indeed 
sensitive to the NN interaction at high medium densities ($\rho\gtrsim\rho_0$ 
in the overlap of the $^3$He projectile with the target nucleus), we have done 
the CC analysis of the \he3t scattering using the \cx form factors (\ref{g7}) 
given by different choices of the IV density dependence of the CDM3Y6 
interaction, in the same way as done with the \pn scattering. The CC results 
(see, e.g., Fig.~\ref{f12}) for the \he3t scattering show that the soft IV 
density dependence of the CDM3Y6 interaction gives a much poorer 
description of the rainbow shoulder seen in the \he3t data at large angles. 
As a result, the slope difference of the NM symmetry energy with  
the increasing NM density shown in Fig.~\ref{f7} can also be traced 
in the folding model analysis of the \he3t scattering to the IAS, and the 
best CC fit to the data prefers again the stiff IV density dependence 
of the CDM3Y6 interaction. Taken together with the results of the folding 
model analysis of the \pn scattering to the IAS discussed in Sec.~\ref{sse}, 
we conclude that the EOS of asymmetric NM with a soft density dependence 
of the symmetry energy $S(\rho)$ is not realistic and should not be used 
in the studies of neutron star matter.

\section{Summary}
A consistent folding model study of the \cx \pn and \he3t scattering 
to the IAS of the target has been performed, where the same density- and 
isospin dependent effective NN interaction has been used to calculate 
the isospin dependent optical potentials and \cx form factors for the 
input of the Lane equations (\ref{e5a})-(\ref{e5b}).

To probe the isospin dependence of the NN interaction, a complex 
IV density dependence of the CDM3Y6 interaction has been constructed 
based on the microscopic JLM nucleon OP in the NM limit \cite{Je77}. 
Such an IV density dependence was used with the IS density 
dependence of the CDM3Y6 interaction parametrized earlier in 
Ref.~\cite{Kho97}, and tested later on in numerous folding model 
studies of nuclear scattering. 
The CC analysis of both the \pn and \he3t scattering using the folded
OP and \cx FF has shown that the (real) IV density dependence of the 
CDM3Y6 interaction (based on the JLM results) needs to be enhanced 
by about $30\sim 40\%$ to give a consistently good CC description 
of the \pn and \he3t data.

The JLM-based IV density dependence of the (real) CDM3Y6 interaction 
has been used also in the HF calculation of asymmetric NM, where the 
nuclear symmetry energy $S(\rho)$ could be obtained within the range 
of empirical values only if the IV density dependence is scaled 
by a factor $N_{V1}\approx 1.3-1.6$, in a close agreement with the 
results of the folding model analysis of the \cx scattering.  

The density dependence of the NM symmetry energy has been further
probed in a direct test of the CDM3Y6 interaction, where we obtained
a consistently worse CC description of the \pn and \he3t data if a 
soft IV density dependence is used instead of the stiff (JLM-based) 
one. Such an assumption has been made recently for the EOS of the 
uniform, $\beta$-stable core of neutron star \cite{Loan11}, and 
a significant reduction of the gravitational mass $M$ and radius $R$ 
of neutron star (away from the empirical boundaries) was 
found when the density dependence of the NM symmetry energy is changed 
from the stiff to the soft behavior. The results of these two 
complimentary studies not only allow us to make a more definitive 
conclusion about the slope of the NM symmetry energy but also stress 
the importance of the experiments on the \cx scattering to the IAS for
the study of the isospin dependence of the \nA and \AA OP. It would be 
of great interest, therefore, to have such experiments pursued 
at the modern rare isotope beam facilities. 

\section*{Acknowledgement}
The present research has been supported by the National Foundation 
for Scientific and Technological Development (NAFOSTED) under 
Project Nr. 103.04-2011.21. We also thank Marcella Grasso
and Eduardo Garrido for providing the microscopic nuclear densities 
for the folding model calculation. 

\end{document}